\documentclass[twocolumn]{IEEEtran}
\usepackage[T1]{fontenc}
\usepackage[latin9]{inputenc}
\usepackage{geometry}
\usepackage{textcomp}
\usepackage{amsmath}
\usepackage{amsthm}
\usepackage{amssymb}
\usepackage{stackrel}
\usepackage{graphicx}
\usepackage[unicode=true,
 bookmarks=true,bookmarksnumbered=true,bookmarksopen=true,bookmarksopenlevel=1,
 breaklinks=false,pdfborder={0 0 0},pdfborderstyle={},backref=false,colorlinks=false]
 {hyperref}
\hypersetup{pdftitle={Your Title},
 pdfauthor={Your Name},
 pdfpagelayout=OneColumn, pdfnewwindow=true, pdfstartview=XYZ, plainpages=false}

\makeatletter
\theoremstyle{plain}
\newtheorem{lem}{\protect\lemmaname}
\theoremstyle{definition}
\newtheorem{assumption}{Assumption}
\theoremstyle{remark}
\newtheorem{rem}{\protect\remarkname}
\theoremstyle{plain}
\newtheorem{thm}{\protect\theoremname}

\usepackage[caption=false,font=footnotesize]{subfig}
\usepackage{cite}
\usepackage[caption=false,font=footnotesize]{subfig}
\IEEEoverridecommandlockouts

\makeatother

\providecommand{\lemmaname}{Lemma}
\providecommand{\remarkname}{Remark}
\providecommand{\theoremname}{Theorem}

\begin{document}
\title{Lyapunov-based Adaptive Transformer (LyAT) for Control of Stochastic
Nonlinear Systems}
\author{Saiedeh Akbari$^{*}$, Xuehui Shen$^{*}$, Wenqian Xue$^{*}$, Jordan
C. Insinger$^{*}$, and Warren E. Dixon$^{*}$ \thanks{$^{*}$Department of Mechanical and Aerospace Engineering, University
of Florida, USA Email: \{akbaris, xuehuishen, w.xue, jordan.insinger,
wdixon\}@ufl.edu.}\thanks{This research is supported in part by AFRL project FA8651-24-1-0018,
AFOSR grant FA9550-19-1-0169, and AFOSR grant FA9550-21-1-0157. Any
opinions, findings, and conclusions or recommendations expressed in
this material are those of the author(s) and do not necessarily reflect
the views of the sponsoring agencies.}}
\maketitle
\begin{abstract}
This paper presents a novel Lyapunov-based Adaptive Transformer (LyAT)
controller for stochastic nonlinear systems. While transformers have
shown promise in various control applications due to sequential modeling
through self-attention mechanisms, they have not been used within
adaptive control architectures that provide stability guarantees.
Existing transformer-based approaches for control rely on offline
training with fixed weights, resulting in open-loop implementations
that lack real-time adaptation capabilities and stability assurances.
To address these limitations, a continuous LyAT controller is developed
that adaptively estimates drift and diffusion uncertainties in stochastic
dynamical systems without requiring offline pre-training. A key innovation
is the analytically derived adaptation law constructed from a Lyapunov-based
stability analysis, which enables real-time weight updates while guaranteeing
probabilistic uniform ultimate boundedness of tracking and parameter
estimation errors. Experimental validation on a quadrotor demonstrates
the performance of the developed controller.
\end{abstract}

\begin{IEEEkeywords}
transformers, neural networks, Lyapunov methods, adaptive control,
nonlinear control systems
\global\long\def\SS{\mathbb{S}}%
\global\long\def\RR{\mathbb{R}}%
\global\long\def\EE{\mathbb{E}}%
\global\long\def\nz{\left\Vert z\right\Vert }%
\global\long\def\ne{\left\Vert e\right\Vert }%
\global\long\def\nt{\left\Vert \widetilde{\theta}\right\Vert }%
\global\long\def\nzz{\left\Vert z\right\Vert ^{2}}%
\global\long\def\nee{\left\Vert e\right\Vert ^{2}}%
\global\long\def\ntt{\left\Vert \widetilde{\theta}\right\Vert ^{2}}%
\global\long\def\tq{\triangleq}%
\global\long\def\Linf{\mathcal{L}_{\infty}}%
\global\long\def\yy{\mathcal{Y}}%
\global\long\def\uu{\mathcal{U}}%
\global\long\def\grad{\mathcal{\nabla_{\widetilde{\theta}}}}%
\global\long\def\tvec{\mathcal{\text{vec}}}%
\global\long\def\sgn{\mathcal{\text{sgn}}}%
\global\long\def\nsig{\left\Vert \Sigma\Sigma^{\top}\right\Vert _{\infty}}%
\global\long\def\rr{\mathbb{R}}%
\end{IEEEkeywords}

\section{Introduction}

Deep learning has emerged as a powerful tool for nonlinear control
due to its universal function approximation capabilities \cite{Anderson1995},
and has been used as means of identification and control for dynamics
with unstructured uncertainties \cite{OConnell.Shi.ea2022,Shi.Shi.ea2019,Punjani.Abbeel2015,Bansal.Akametalu.ea2016,Li.Qian.ea2017,Zhou.Helwa.ea2017,Abbeel.Coates.ea2010}.
Transformers are deep learning architectures that have revolutionized
sequential modeling through self-attention mechanisms, making them
well-suited for control problems requiring long-term temporal reasoning
\cite{Vaswani.2017}. Transformers offer significant advantages for
controlling stochastic nonlinear systems due to their ability to handle
uncertainty through attention mechanisms and efficiently process complex
sequential data in parallel. Unlike traditional neural network architectures,
transformers allow for simultaneous processing of entire sequences
and therefore provide global context awareness, which is crucial in
control applications where past states significantly influence current
control decisions.

The architectural evolution from recurrent neural networks (RNNs)
to long short-term memory (LSTM) networks to transformers indicates
the progression in modeling temporal dependencies (e.g., \cite{Medsker.Jain1999}
and \cite{Hochreiter.Schmidhuber.1997}). RNNs suffer from vanishing
gradients that limit long-term dependency capture \cite{Salehinejad.Sankar.ea2017},
while LSTMs improve retention through gating mechanisms but compress
history into fixed-size hidden states, potentially losing critical
information \cite{Yu.Si.ea2019}. Transformers overcome these limitations
through attention mechanisms that maintain direct access to all previous
states and dynamically weight historical information based on system
performance without computationally expensive recurrent loops \cite{Vaswani.2017}. 

Transformers are the primary architecture used in natural language
processing (NLP) in the recent years, but their application to control
has been limited. Since their introduction \cite{Vaswani.2017}, there
have been some efforts to apply transformers in control systems, including
PID controller integration \cite{Nguyen.Uribe.ea2024}, reinforcement
learning \cite{Kim.Lee.ea2025}, and specialized applications such
as quantum feedback control and power flow adjustment (cf., \cite{Chen.Luo.ea2025}
and \cite{Vaidhyanathan.Marquardt.ea2024}). However, to the best
of the authors' knowledge, transformers have not been formalized and
integrated within an adaptive control architecture.

Most existing deep learning approaches for control rely on offline
training using sampled input-output datasets (e.g., \cite[Sec. 6.6]{Brunton.Kutz2019}
and \cite{Noda.Arie.ea2014,Sarikaya.Corso.ea2017,Nguyen.Cheah2022}).
These architectures are typically applied as feedforward terms with
fixed weights that cannot be updated in real-time. Recent breakthroughs
develop Lyapunov-based update laws that adjust neural network weights
in real-time through analytical update laws constructed from a stability
analysis \cite{Patil.Le.ea2022,Patil.Le.ea.2022,Griffis.Patil.ea.2023,Akbari.Griffis.ea2023,Akbari.Nino.ea2024,Akbari.Patil.ea2025,Shen.Griffis.eatoappear}.
Specifically, the weights are updated using analytically-derived laws
designed through Lyapunov-based stability analysis that enables real-time
adaptation of the parameters of the architecture without requiring
offline pre-training \cite{Patil.Le.ea2022,Patil.Le.ea.2022}. While
Lyapunov-based Deep RNNs, LSTMs, and Deep LSTMs have been developed
for deterministic systems, despite the superior ability of transformers
to capture long-term dependencies through attention mechanisms, they
have not been integrated within adaptive control frameworks with stability
guarantees.

The control of stochastic systems presents additional analytical challenges
but offers considerable advantages since modeling uncertainty as stochastic
processes makes control designs less conservative than assuming worst-case
bounds \cite{Kumar.Varaiya2015}. Neural network-based learning methods
are particularly valuable for nonlinear stochastic systems due to
their black-box approximation capabilities \cite{Li.Chen.ea2009,Chen.Liu.ea2013,Li.Li.ea2018,Wang.Liu.ea2019a,Wang.Chen2020,Wang.You.ea2022,Chen.Jiao.ea2009}.
However, despite considerable performance improvements from Lyapunov-based
learning methods, their application to deep learning architectures
for stochastic systems remains very limited (cf., \cite{Akbari.Nino.ea2024}
and \cite{Chen.Mei.ea2024}).

Addressing these gaps, this work develops the first Lyapunov-based
Adaptive Transformer (LyAT) controller for stochastic nonlinear systems.
Adapting transformers from NLP applications to real-time control requires
overcoming substantial architectural and theoretical challenges. The
key contributions of the developed result include the following.:
\begin{enumerate}
\item Transformer architecture design for real-time adaptive control: Unlike
NLP applications where inputs (tokens) are straightforward, control
applications require strategic input selection to assist with control
objectives and system stabilization. In this paper, where the objective
is tracking a desired trajectory, the inputs are designed as states,
desired trajectories, and tracking errors to enable the attention
mechanism to focus on tracking performance. The architecture is modified
by removing embedding layers and modifying projection matrices to
accommodate asymmetric encoder-decoder dimensions. The encoder processes
rich state and trajectory information while the decoder focuses on
uncertainty estimation (drift and diffusion), with the cross-attention
mechanism handling the dimension mismatch through its projection matrices.
This new asymmetric design is particularly beneficial for control
applications because it eliminates the constraint that encoder and
decoder inputs must be the same size, allowing each to be optimized
for its specific role. The resulting continuous LyAT controller adaptively
estimates drift and diffusion uncertainties which are then used in
the control law.
\item Lyapunov-based parameter adaptation without pre-training: Existing
transformers require massive offline training and training data which
is mostly not available or feasible to attain in control applications.
In this paper, an analytical parameter update law is constructed from
Lyapunov-based stability analysis enabling real-time weight adjustment
across all transformer layers using only streaming system feedback.
This eliminates offline training while providing provable performance
certificates.
\item Unified architecture and stability analysis: A constructive Lyapunov-based
analysis guarantees probabilistic uniform ultimate boundedness of
tracking and parameter estimation errors. This work streamlines the
stability analysis approach of \cite{Akbari.Nino.ea2024} by employing
a single neural network architecture to simultaneously compensate
for all uncertain terms arising from both drift and diffusion dynamics.
In contrast to the more traditional feedforward DNN architecture in
the prior work in \cite{Akbari.Nino.ea2024} that required multiple
DNNs for separate uncertainty components, in addition to the unique
adaptive transformer design, the developed method in this paper also
includes a unified network that incorporates the tracking error into
its input, rather than solely the system states. This design enables
efficient adaptation across all uncertainty channels while reducing
computational and memory overhead.
\item Experimental results: Experimental validation on a Freefly Astro quadrotor
tracking a figure-8 trajectory demonstrates rapid convergence from
initial conditions with a root mean square (RMS) tracking error of
0.2175 meters over 240 seconds of flight, demonstrating efficacy of
the theoretical stability guarantees and practical applicability of
the developed approach.
\end{enumerate}

\section{Problem Formulation}

\subsection{Notation}

An identity matrix of size $n$ is denoted by $I_{n}\in\rr^{n\times n}$.
The right pseudo-inverse of full row rank matrix $A\in\rr^{n\times m}$
is denoted by $A^{+}$, where $A^{+}\triangleq A^{\top}\left(AA^{\top}\right)^{-1}$.
A zero matrix with dimension of $p\times q$ is denoted as $\mathbf{0}_{p\times q}$.
The�rectified�linear�unit (ReLU)�activation�function is denoted as
$\sigma_{\text{ReLU}}\left(x\right)\tq\max\left(0,x\right)$. Given
some functions $f$ and $g$ and some $w\in\mathbb{R}$, the notation
$f(w)=\mathcal{O}^{m}(g(w))$ means that $\left\Vert f(w)\right\Vert \leq M\left\Vert g(w)\right\Vert ^{m}$
for all $w\geq w_{0}$, where $M\in\mathbb{R}_{>0}$ and $w_{0}\in\mathbb{R}$
denote constants. The vectorization operator is denoted by $\mathrm{vec}\left(\cdot\right)$,
i.e., given $A\triangleq\left[a_{i,j}\right]\in\mathbb{R}^{n\times m}$,
$\mathrm{vec}\left(A\right)\triangleq\left[a_{1,1},\ldots,a_{n,1},\ldots,a_{1,m},\ldots,a_{n,m}\right]{}^{\top}$.
Given input $z\in\rr^{k}$ and tunable constant parameters $\gamma,\beta\in\rr$,
the layer norm function LayerNorm$\left(z,\gamma,\beta\right)\in\rr^{k}$
is defined as $\text{LayerNorm}\left(z,\gamma,\beta\right)\triangleq\gamma\frac{\left(z-\mu_{z}\right)}{\sigma_{z}}+\beta,$
where $\mu_{z}\triangleq\frac{1}{k}\stackrel[i=1]{k}{\sum}z_{i}$
and $\sigma_{z}\triangleq\sqrt{\frac{1}{k}\stackrel[i=1]{k}{\sum}\left(z_{i}-\mu_{z}\right)^{2}}$.
Given an input $Z\tq\left[z_{1},\cdots,z_{q}\right]^{\top}\in\RR^{q\times k}$,
the LayerNorm is applied row-wise, that is, the $i^{\text{th}}$ row
$z_{i}\in\RR^{k}$ is normalized independently as LayerNorm$\left(z_{i},\gamma,\beta\right)$,
resulting in an output LayerNorm$\left(Z,\gamma,\beta\right)\in\RR^{q\times k}$.
Given an input $z\tq\left\{ {\tt z}_{1},\cdots,{\tt z}_{k}\right\} \in\rr^{k}$,
the softmax function is defined as softmax$\left(z\right)\tq\frac{\exp\left({\tt z}_{j}\right)}{\sum_{j=1}^{k}\exp\left({\tt z}_{j}\right)}$.
Given the input $Z$, the softmax function is applied row-wise, that
is, the $i^{\text{th}}$ row $z_{i}\in\RR^{k}$ is independently normalized
as softmax$\left(z_{i}\right)$, resulting in an output matrix softmax$\left(Z\right)\in\RR^{k\times q}$.
For a square matrix $A\in\RR^{n\times n}$, the trace operator is
defined as $\text{tr}\left(A\right)=\stackrel[i=1]{n}{\sum}a_{i,i}$,
where $a_{i,i}$ represents the element of the $i^{\text{th}}$ row
on the $i^{\text{th}}$ column. From \cite[Chapter 2, Eq. 13]{Magnus.Neudecker2019},
the trace to vector property 
\begin{equation}
\text{tr}\left(A^{\top}B\right)=\tvec\left(A\right)^{\top}\tvec\left(B\right)\label{eq: trace to vec}
\end{equation}
holds for matrices $A$ and $B$. From \cite[Chapter 1, Eq. 25]{Magnus.Neudecker2019},
\begin{equation}
\text{tr}\left(ABC\right)=\text{tr}\left(BCA\right)=\text{tr}\left(CAB\right).\label{eq: order of multiplication property}
\end{equation}
Additionally, if $A$ and $B$ are positive semi-definite matrices,
then 
\begin{equation}
\text{tr}\left(AB\right)\leq\text{tr}\left(A\right)\text{tr}\left(B\right).\label{eq: separating trace inequality}
\end{equation}
Given a function $h:\RR^{n}\to\RR^{n}$, the notation $\underset{a\to b^{-}}{\lim}h\left(a\right)$
denotes the left-hand limit of $h$ at $b$. The $p$-norm is denoted
by $\left\Vert \cdot\right\Vert _{p}$, where the subscript is suppressed
when $p=2$. The Frobenius norm is denoted by $\left\Vert \cdot\right\Vert _{F}\triangleq\left\Vert \mathrm{vec}(\cdot)\right\Vert $.
For a bounded function $f:\RR_{\geq0}\to\RR^{n\times m}$, $\left\Vert f\right\Vert _{F\infty}\tq\underset{t\in\RR_{\geq0}}{\sup}\left\Vert f\right\Vert _{F}$.
The space of $k$-times differentiable functions is denoted by $\mathcal{C}^{k}$,
and a $\mathcal{C}^{\infty}$-smooth function is an infinitely differentiable
function. In the filtered probability space of $\left({\bf \Omega},\,\mathbb{F},\,\mathbb{F}_{t},{\rm P}\right)$,
${\bf \Omega}$ represents the event space, $\mathbb{F}$ denotes
a $\sigma$-algebra of the subsets of ${\bf \Omega}$ and represents
the event space, $\mathbb{F}_{t}$ is a complete filtration given
by the family of $\sigma$-algebras up to time $t$, i.e., $\mathbb{F}_{S}:\mathbb{F}_{S}\subseteq\mathbb{F}_{t}\hspace{1em}\forall t\in\left[0,t\right]$,
and ${\rm P}$ is a probability measure, where the filtration is complete
in the sense that it includes all events with probability measure
zero (see \cite{Lanchares.Haddad2023}). Consider a probability space
of $\left({\bf \Omega},\,\mathbb{F},\,{\rm P}\right)$. Therefore,
for any events $A,\,B\in\mathbb{F}$ such that $A\subseteq B$, the
monotonicity property states that \cite[eq. 2.5]{Billingsley2017}
\begin{equation}
{\rm P}\left(A\right)\leq{\rm P}\left(B\right).\label{eq:monotonicity}
\end{equation}
Consider a stochastic differential equation (SDE) as ${\rm d}x={\bf f}\left(x\right){\rm d}t+{\bf g}\left(x,t\right){\rm d}\omega$.
Then, for some function $V\in\mathcal{C}^{2}$ associated with this
SDE, let the infinitesimal generator $\mathcal{L}$ of the function
$V\left(x\right)$ be defined as \cite[eq. 4.12]{Kushner1967}
\begin{gather}
\mathcal{L}V\tq\frac{\partial V}{\partial x}{\bf f}\left(x\right)+\frac{1}{2}\text{tr}\left({\bf g}\left(x,t\right)^{\top}\frac{\partial^{2}V}{\partial x^{2}}{\bf g}\left(x,t\right)\right).\label{eq: LV_L-1}
\end{gather}

\begin{lem}
\label{thm:probability}\textup{(\cite[Lemma 1]{Akbari.Nino.ea2024})
For the Ito process ${\tt z}\in\RR^{n}$ and function ${\tt V}$,
assume }\\
(A1) \label{(A1)--is}${\tt V}$ is non-negative, ${\tt V}\left(0\right)=0$,
and ${\tt V}\in\mathcal{C}^{2}$ over the open and connected set $Q_{m}\tq\left\{ {\tt z}:{\tt V}\left({\tt z}\right)<m\right\} $,
where $m\in\RR_{>0}$ is a bounding constant,\\
(A2) \label{(A2)--is}${\tt z}\left(t\right)$ is a continuous strong
Markov process defined until at least some $\tau^{\prime}>\tau_{m}=\inf\left\{ t:{\tt z}\left(t\right)\notin Q_{m}\right\} $
with probability one.\footnote{This assumption guarantees the existence of the process up to $\tau^{\prime}$
with probability one.}\\
If $\mathcal{L}{\tt V}\left({\tt z}\right)\leq-{\tt k}_{1}{\tt V}\left({\tt z}\right)+{\tt k}_{2}$
in $Q_{m}$ for ${\tt k}_{1},{\tt k}_{2}>0$, then for $\ell\leq m$,
${\tt z}\left(t\right)$ is uniformly ultimately bounded in probability
(UUB-p) with the probability
\begin{gather}
{\rm P}\left(\underset{t\leq s<\infty}{\sup}{\tt V}\left({\tt z}\left(s\right)\right)\geq\ell\right)\leq\frac{1}{m}{\tt V}\left({\tt z}\left(0\right)\right)\nonumber \\
+\frac{1}{\ell}{\tt V}\left({\tt z}\left(0\right)\right)\exp\left(-{\tt k}_{1}t\right)+\frac{{\tt k}_{2}}{{\tt k}_{1}\ell}.\label{eq: probability lemma main result}
\end{gather}
\end{lem}

\subsection{System Dynamics and Control Objective}

Consider the system dynamics modeled as 
\begin{equation}
{\rm d}x=\left(f\left(x\right)+g_{1}\left(x\right)u\left(t\right)\right){\rm d}t+g_{2}\left(x\right)\Sigma\left(t\right){\rm d}\omega,\label{eq: dynamics}
\end{equation}
where $x:\mathbb{R}_{\ge0}\rightarrow\mathbb{R}^{n}$ denotes the
state, $u:\rr_{\geq0}\rightarrow\rr^{m}$ denotes the control input,
$f:\rr^{n}\rightarrow\rr^{n}$ denotes an unknown continuous drift
vector, and $g_{1}:\rr^{n}\rightarrow\rr^{n\times m}$ denotes the
known full-row rank and bounded control effectiveness matrix.\footnote{The control development in \cite{Sun.Greene.ea2021} can be used to
account for an uncertain, linearly parametrizable control effectiveness
matrix $g$.} Additionally, in \eqref{eq: dynamics}, $g_{2}:\RR^{n}\to\RR^{n\times s}$
denotes the unknown continuous diffusion matrix, $\Sigma:\RR_{\geq0}\to\RR^{s\times s}$
denotes the symmetric Borel-measurable covariance matrix, and $\omega\in\RR^{s}$
denotes the $s$-dimensional independent standard Wiener process defined
on the complete filtered probability space $\left({\bf \Omega},\,\mathbb{F},\,\mathbb{F}_{t},{\rm P}\right)$. 

The control objective is to design a LyAT-based controller that ensures
the states track a user-defined desired trajectory, $x_{d}:\mathbb{R}_{\ge0}\rightarrow\mathbb{R}^{n}$.
To quantify the tracking objective, the tracking error, $e:\mathbb{R}_{\geq0}\rightarrow\mathbb{R}^{n}$,
is defined as 
\begin{align}
e & \triangleq x-x_{d}.\label{eq: Tracking error}
\end{align}

\begin{assumption}
The desired trajectory $x_{d}$ is designed to be sufficiently smooth,
i.e., $\left\Vert x_{d}\left(t\right)\right\Vert \le\overline{x_{d}}$
and $\left\Vert \dot{x}_{d}\left(t\right)\right\Vert \le\overline{\dot{x}_{d}}$
for all $t\in\mathbb{R}_{\ge0}$, where $\overline{x_{d}},\overline{\dot{x}_{d}}\in\mathbb{R}_{>0}$
are known constants. 
\end{assumption}
To adapt to the uncertainties in the diffusion matrix in the subsequent
stability analysis, Taylor's expansion theorem is applied to the vectorized
diffusion matrix $g_{2}$, yielding
\begin{align}
\tvec\left(g_{2}\left(x\right)\right) & =\text{vec}\left(g_{2}\left(x_{d}\right)\right)+\frac{\partial\text{vec}\left(g_{2}\left(x_{d}\right)\right)}{\partial x_{d}}\left(x-x_{d}\right)\nonumber \\
 & \quad+\mathcal{O}\left(\left\Vert x-x_{d}\right\Vert ^{2}\right)\nonumber \\
 & =\text{vec}\left(g_{2}\left(x_{d}\right)\right)+\frac{\partial\text{vec}\left(g_{2}\left(x_{d}\right)\right)}{\partial x_{d}}e+\mathcal{O}\left(\left\Vert e\right\Vert ^{2}\right)\nonumber \\
 & =G_{2}\left(e,x_{d}\right)e+\tvec\left(g_{2}\left(x_{d}\right)\right),\label{eq: vectorized g2}
\end{align}
where $G_{2}:\RR^{n}\times\RR^{n}\to\RR^{ns\times n}$ is a $\mathcal{C}^{\infty}$
function, and $\tvec\left(g_{2}\left(x_{d}\right)\right)$ is upper-bounded
as $\left\Vert \tvec\left(g_{2}\left(x_{d}\right)\right)\right\Vert \leq\overline{g}$,
where $\bar{g}\in\mathbb{R}_{>0}$ is unknown.

\section{Transformer Neural Network Architecture}

\begin{figure}[tbh]
\begin{centering}
\includegraphics[viewport=20bp 470bp 800bp 1015bp,clip,width=0.98\columnwidth]{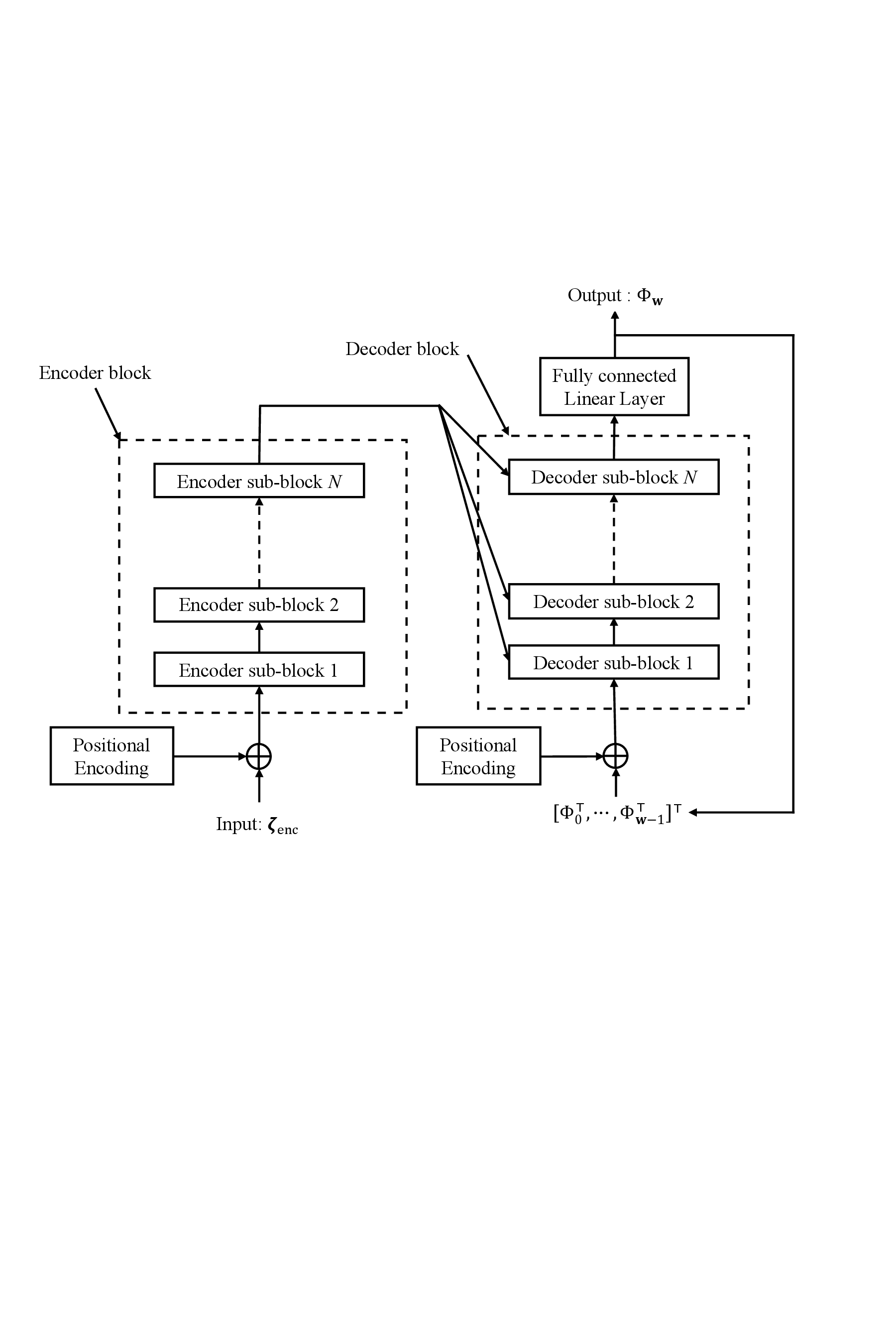}
\par\end{centering}
\caption{\label{fig:Transformer-Model}An overview of the encoder-decoder transformer
architecture. The transformer architecture consists of $N$ encoder
layers and $N$ decoder layers.}
\end{figure}
\begin{figure}[tbh]
\begin{centering}
\includegraphics[viewport=110bp 520bp 875bp 1150bp,clip,width=0.98\columnwidth]{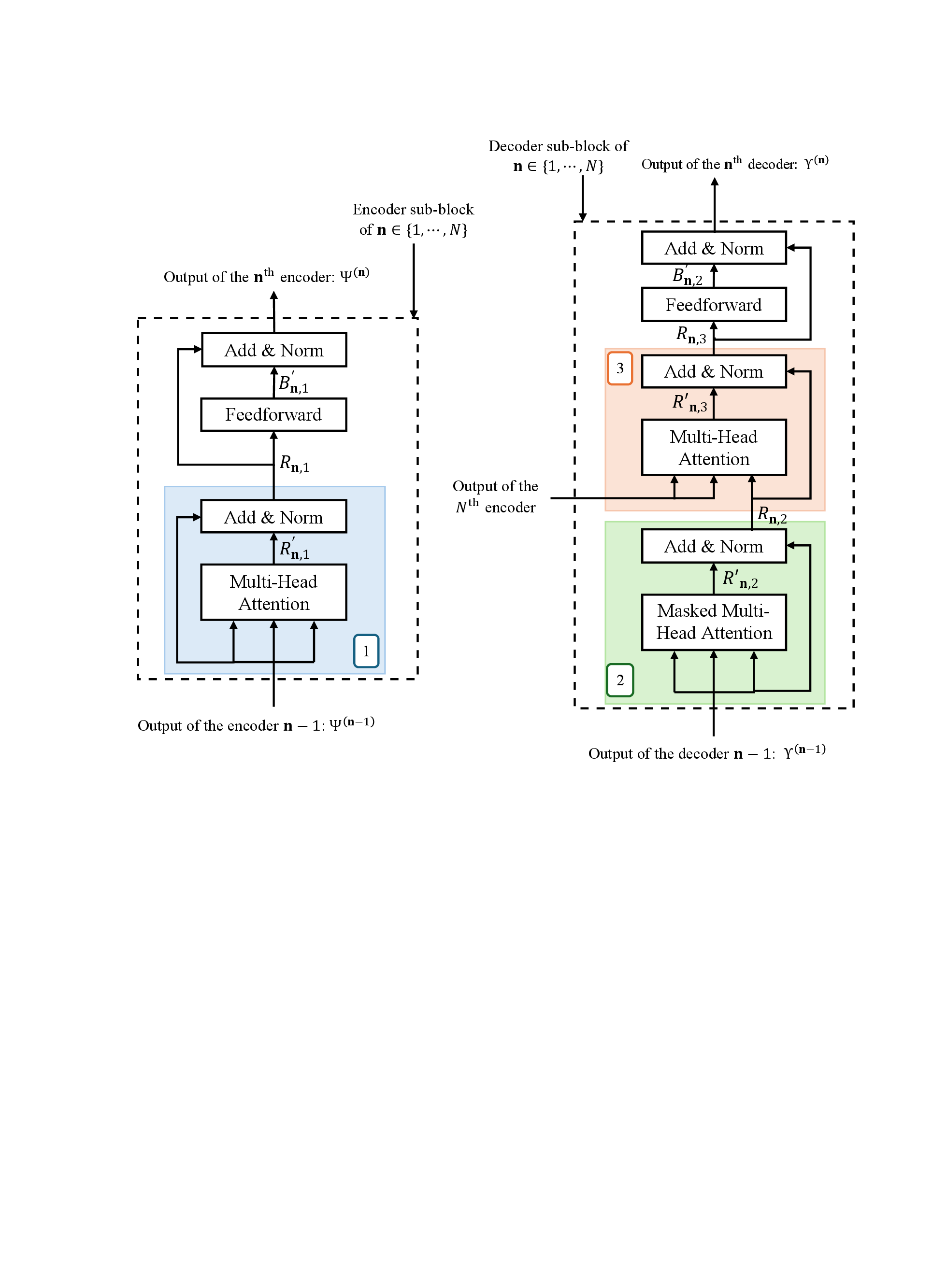}
\par\end{centering}
\caption{\label{fig:EncoderDecoder}A modular schematic of each attention-based
encoder and decoder in a transformer architecture. Across these blocks,
there are three key attention mechanisms, illustrated as 1, 2, and
3: self-attention, masked self-attention, and cross-attention, respectively. }
\end{figure}

A LyAT, denoted as $\Phi:\RR^{3n\boldsymbol{\tau}}\times\RR^{p}\to\RR^{n}$,
is developed to compensate for the drift and diffusion uncertainties,
$\mathcal{F}:\RR^{3n}\to\RR^{n}$, that appear in the subsequent stability
analysis as $\mathcal{F}\left({\tt X}\right)\tq f\left(x\right)+\frac{1}{2}e\left\Vert \Sigma\Sigma^{\top}\right\Vert _{F\infty}\text{tr}\left\{ G_{2}^{\top}\left(e,x_{d}\right)G_{2}\left(e,x_{d}\right)\right\} +\left\Vert \Sigma\Sigma^{\top}\right\Vert _{F\infty}G_{2}^{\top}\left(e,x_{d}\right)\text{vec}\left(g_{2}\left(x_{d}\right)\right)$,
where ${\tt X}\tq\left[x^{\top},x_{d}^{\top},e^{\top}\right]^{\top}\in\RR^{3n}$
is the current state information. The transformer has an encoder-decoder
structure as illustrated in Figures \ref{fig:Transformer-Model} and
\ref{fig:EncoderDecoder}. To form the network's input, the data is
structured into sequences. The input to the encoder is a sequence
of the last $\boldsymbol{\tau}\in\mathbb{Z}_{>0}$ steps, i.e., $\boldsymbol{\zeta}_{\text{enc}}\triangleq\left[x_{1}^{\top},x_{d,1}^{\top},e_{1}^{\top},\cdots,x_{\boldsymbol{\tau}}^{\top},x_{d,\boldsymbol{\tau}}^{\top},e_{\boldsymbol{\tau}}^{\top}\right]^{\top}\in\mathbb{R}^{3n\boldsymbol{\tau}}$.
The input to the decoder is the last $\boldsymbol{\tau}$ estimates
of the LyAT output, $\left[\Phi_{0}^{\top},\cdots,\Phi_{\boldsymbol{\tau}-1}^{\top}\right]^{\top}\in\RR^{n\boldsymbol{\tau}}$.\footnote{The input to the encoder can be viewed as a window of ${\tt X}_{i}$,
for all $i\in\left\{ 1,2,\cdots,\boldsymbol{\tau}\right\} $, as $\boldsymbol{\zeta}_{\text{enc}}=\left[{\tt X}_{1}^{\top},\cdots,{\tt X}_{\boldsymbol{\tau}}^{\top}\right]^{\top}$,
where ${\tt X}_{\boldsymbol{\tau}}$ for notational simplicity is
written as ${\tt X}_{\boldsymbol{\tau}}={\tt X}$.}

\begin{rem}
During the initial transient phase before $\boldsymbol{\tau}$ historical
values are available, the encoder and decoder inputs are populated
with initialization values (e.g., zeros, inputs from offline training
if available, or random inputs) that are sequentially updated with
the actual state measurements and network outputs.
\end{rem}
\begin{rem}
The encoder input $\boldsymbol{\zeta}_{\text{enc}}$, while mathematically
represented as a concatenated vector, is structured and processed
as a sequence of $\boldsymbol{\tau}$ positions for the transformer
architecture. Specifically, each position $i\in\left\{ 1,2,\cdots,\boldsymbol{\tau}\right\} $
contains the concatenated state information $\left[x_{i}^{\top},x_{d,i}^{\top},e_{i}^{\top}\right]^{\top}\in\RR^{3n}$,
enabling the multi-head attention mechanism to capture temporal dependencies
across the historical window.
\end{rem}
\begin{rem}
The encoder focuses on tracking performance by processing a window
of size $\boldsymbol{\tau}$ containing historical state information
$\boldsymbol{\zeta}_{\text{enc}}$ to capture temporal patterns in
system dynamics and tracking behavior. The decoder focuses on function
approximation by receiving a window of the previous $\boldsymbol{\tau}$
transformer output estimates, $\Phi_{0},\cdots,\Phi_{\boldsymbol{\tau}-1}$,
and using cross-attention to query relevant patterns from the encoder.
Through cross-attention, the decoder queries the encoder's tracking-oriented
representations to refine its uncertainty estimates. This architecture
enables the transformer to condition its drift and diffusion compensation
on both historical tracking performance and its own previous approximations.
\end{rem}
The encoder block, consisting of $N$ identical encoder sub-blocks,
maps the input sequence to a sequence of continuous representations,
$\Psi^{\left(N\right)}\in\RR^{3n\boldsymbol{\tau}}$. The latent sequence
$\Psi^{\left(N\right)}$ is then utilized by the decoder stack's cross-attention
mechanism, also consisting of $N$ identical decoder sub-blocks. The
decoder also processes the feedback sequence $\left[\Phi_{0}^{\top},\cdots,\Phi_{\boldsymbol{\tau}-1}^{\top}\right]^{\top}\in\RR^{n\boldsymbol{\tau}}$
and maps that to the final output of the $N$-layer decoder, $\Upsilon^{\left(N\right)}\in\RR^{n\boldsymbol{\tau}}$,
which is then used to compute the transformer estimate in continuous
time. Each encoder and decoder block internally uses multi-head attention
mechanisms and position-wise feedforward networks.
\begin{rem}
In the traditional Transformer architecture (cf., \cite{Vaswani.2017}),
both the input to the multi-head attention in the encoder block and
the input to the masked multi-head attention are first processed through
an input embedding layer. This embedding mechanism is essential in
language models where the inputs are typically tokens or words that
need to be converted to numerical signals. However, in this paper,
the LyAT is applied to estimate the drift and diffusion uncertainties
of a stochastic dynamical system. Since the inputs in this context
are already numerical values, the input embedding layer is not implemented.
\end{rem}

\subsection{Positional Encoding}

The encoder and decoder blocks both contain a positional encoding
to provide a sense of temporal or spatial order to the input data.
Therefore, the inputs of the encoder and the decoder blocks are first
fed into a positional encoding mechanism. For the encoder, the positional
encoding for each $i\in\left\{ 1,\cdots,\boldsymbol{\tau}\right\} $
is defined as \cite{Vaswani.2017}
\begin{align}
{\rm PE}_{\left(i-1,2j\right)}^{\text{enc}} & =\sin\left(\frac{i-1}{10000^{\frac{2j}{3n}}}\right),\nonumber \\
{\rm PE}_{\left(i-1,2j+1\right)}^{\text{enc}} & =\cos\left(\frac{i-1}{10000^{\frac{2j}{3n}}}\right),\label{eq: positional encoding - encoder}
\end{align}
where $j\in\left\{ 0,1,\cdots,\frac{3n-1}{2}\right\} $. For each
$i\in\left\{ 1,\cdots,\boldsymbol{\tau}\right\} $ in the sequence,
positional encoding vectors of the encoder ${\rm PE}_{i-1}^{\text{enc}}\triangleq\left[{\rm PE}_{\left(i-1,0\right)}^{\text{enc}},{\rm PE}_{\left(i-1,1\right)}^{\text{enc}},\cdots,{\rm PE}_{\left(i-1,3n-1\right)}^{\text{enc}}\right]^{\top}$are
stacked as $\mho_{\text{enc}}\triangleq\left[{\rm PE}_{0}^{\text{enc}\top},{\rm PE}_{1}^{\text{enc}\top},\cdots,{\rm PE}_{\boldsymbol{\tau}-1}^{\text{enc}\top}\right]^{\top}\in\RR^{3n\boldsymbol{\tau}}$.
For the decoder, the positional encoding for each $i\in\left\{ 1,\cdots,\boldsymbol{\tau}\right\} $
is defined as
\begin{align}
{\rm PE}_{\left(i-1,2j\right)}^{\text{dec}} & =\sin\left(\frac{i-1}{10000^{\frac{2j}{n}}}\right),\nonumber \\
{\rm PE}_{\left(i-1,2j+1\right)}^{\text{dec}} & =\cos\left(\frac{i-1}{10000^{\frac{2j}{n}}}\right),\label{eq: positional encoding - decoder}
\end{align}
where $j\in\left\{ 0,1,\cdots,\frac{n-1}{2}\right\} $. For each $i\in\left\{ 0,\cdots,\boldsymbol{\tau}-1\right\} $
in the sequence, positional encoding vectors of the decoder ${\rm PE}_{i-1}^{\text{dec}}\triangleq\left[{\rm PE}_{\left(i-1,0\right)}^{\text{dec}},{\rm PE}_{\left(i-1,1\right)}^{\text{dec}},\cdots,{\rm PE}_{\left(i-1,n-1\right)}^{\text{dec}}\right]^{\top}$are
stacked as $\mho_{\text{dec}}\triangleq\left[{\rm PE}_{0}^{\text{dec}\top},{\rm PE}_{1}^{\text{dec}\top},\cdots,{\rm PE}_{\boldsymbol{\tau}-1}^{\text{dec}\top}\right]^{\top}\in\RR^{n\boldsymbol{\tau}}$. 

\subsection{Attention Mechanism}

The attention mechanism is designed as a mapping of a query and a
set of key-value pairs to an output, where the query, keys, and values
are all vector outputs of fully-connected layers. The output of the
attention mechanism is computed as a weighted sum of the values. The
weight associated with each value is computed using a compatibility
function of the query with the corresponding key. This weighted sum
is computed over $H\in\mathbb{Z}_{>0}$ heads, resulting in a multi-head
attention mechanism. The multi-head design allows the transformer
to perform attention multiple times in parallel \cite{Vaswani.2017}.
Since each head has its own set of query, keys, and values, the model
is able to focus on different parts of the input simultaneously and
learn different aspects of the data that may not be captured by a
single attention head. A mathematical model of the multihead attention
mechanism is subsequently provided.

An attention mechanism computes its output from two sources of input.
One source, denoted ${\tt A}_{\boldsymbol{{\tt {\tt n}}},\ell}\left(t\right)\triangleq\left[\left({\tt a}_{\ell,1}^{\left(\boldsymbol{{\tt n}}\right)}\right)^{\top},\cdots,\left({\tt a}_{\ell,\boldsymbol{\tau}}^{\left(\boldsymbol{{\tt n}}\right)}\right)^{\top}\right]^{{\bf \top}}$,
provides the input for the key and value vectors, and the other source,
denoted as ${\tt C}_{\boldsymbol{{\tt n}},\ell}\left(t\right)\triangleq\left[\left({\tt c}_{\ell,1}^{\left(\boldsymbol{{\tt n}}\right)}\right)^{\top},\cdots,\left({\tt c}_{\ell,\boldsymbol{\tau}}^{\left(\boldsymbol{{\tt n}}\right)}\right)^{\top}\right]^{\top}$,
provides the input for the query, for $\boldsymbol{{\tt n}}\in\left\{ 1,2,\cdots,N\right\} $
and $\ell\in\left\{ 1,2,3\right\} $.\footnote{For the self-attention mechanism in the encoder (Attention Mechanism
1) of the first layer, ${\tt A}_{1,1}\left(t\right)={\tt C}_{1,1}\left(t\right)=\boldsymbol{\zeta}_{\text{enc}}\left(t\right)+\mho_{\text{enc}}\left(t\right)$.
For the masked multi-head attention mechanism in the decoder (Attention
Mechanism 2) of the first layer, ${\tt A}_{1,2}\left(t\right)={\tt C}_{1,2}\left(t\right)={\tt \left[\Phi_{0}^{\top},\cdots,\Phi_{\boldsymbol{\tau}-1}^{\top}\right]^{\top}}+\mho_{\text{dec}}$.
For the cross-attention mechanism in the decoder (Attention Mechanism
3) of the first layer, ${\tt A}_{1,3}\left(t\right)=\Psi^{\left(N\right)}\left(t\right)$
and ${\tt C}_{1,3}\left(t\right)=R_{1,2}\left(t\right)$.} The output for each of the attention mechanism of each layer, denoted
as $R_{\boldsymbol{{\tt n}},1}\in\RR^{3n\boldsymbol{\tau}},R_{\boldsymbol{{\tt n}},2},R_{\boldsymbol{{\tt n}},3}\in\RR^{n\boldsymbol{\tau}}$,
is defined as 
\begin{equation}
R_{\boldsymbol{{\tt n}},\ell}\left(t\right)\triangleq\text{LayerNorm}\left({\tt C}_{\boldsymbol{{\tt n}},\ell}\left(t\right)+R_{\boldsymbol{{\tt n}},\ell}^{\prime}\left(t\right),\gamma_{\boldsymbol{{\tt n}},\ell},\beta_{\boldsymbol{{\tt n}},\ell}\right),\label{eq: output of attention mech. R}
\end{equation}
for $\ell\in\left\{ 1,2,3\right\} $ and $\boldsymbol{{\tt n}}\in\left\{ 1,2,\cdots,N\right\} $,
where $\gamma_{\boldsymbol{{\tt n}},\ell},\beta_{\boldsymbol{{\tt n}},\ell}\in\mathbb{R}_{>0}$
are user-selected constants. The output of the Multi-Head Attention
for each layer are denoted as $R_{\boldsymbol{{\tt n}},1}^{\prime}\in\RR^{3n\boldsymbol{\tau}},R_{\boldsymbol{{\tt n}},2}^{\prime},R_{\boldsymbol{{\tt n}},3}^{\prime}\in\RR^{n\boldsymbol{\tau}}$
and defined as
\[
R_{\boldsymbol{{\tt n}},\ell}^{\prime}\left(t\right)\triangleq\left(\left[\left({\tt H}_{\ell,1}^{\left(\boldsymbol{{\tt n}}\right)}\right)^{\top},\left({\tt H}_{\ell,2}^{\left(\boldsymbol{{\tt n}}\right)}\right)^{\top},\cdots,\left({\tt H}_{\ell,H}^{\left(\boldsymbol{{\tt n}}\right)}\right)^{\top}\right]\mathcal{W}_{\ell}^{\left(\boldsymbol{{\tt n}}\right)}\right)^{\top},
\]
where $\mathcal{W}_{1}^{\left(\boldsymbol{{\tt n}}\right)}\in\RR^{3n\times3n\boldsymbol{\tau}},\mathcal{W}_{2}^{\left(\boldsymbol{{\tt n}}\right)},\mathcal{W}_{3}^{\left(\boldsymbol{{\tt n}}\right)}\in\RR^{n\times n\boldsymbol{\tau}}$
denote the matrices of multi-head weights. These multi-head matrices
can be vectorized as $\theta_{\mathcal{W}}=$ $\bigg[\text{vec}\left(\mathcal{W}_{1}^{\left(1\right)}\right)^{\top},\cdots,$
$\text{vec}\left(\mathcal{W}_{1}^{\left(N\right)}\right)^{\top},$
$\text{vec}\left(\mathcal{W}_{2}^{\left(1\right)}\right)^{\top},\cdots,$
$\text{vec}\left(\mathcal{W}_{2}^{\left(N\right)}\right)^{\top},$
$\text{vec}\left(\mathcal{W}_{3}^{\left(1\right)}\right)^{\top},\cdots,$
$\text{vec}\left(\mathcal{W}_{3}^{\left(N\right)}\right)^{\top}\bigg]^{\top}$$\in\RR^{11n^{2}N\boldsymbol{\tau}}$.
For each head of the multihead attention block $h\in\left\{ 1,2,\cdots,H\right\} $,
and $\boldsymbol{{\tt n}}\in\left\{ 1,2,\cdots,N\right\} $,

\begin{alignat}{1}
{\tt H}_{\ell,h}^{\left(\boldsymbol{{\tt n}}\right)}\left(t\right) & \triangleq\begin{cases}
\text{softmax}\bigg(\frac{Q_{\ell,h}^{\left(\boldsymbol{{\tt n}}\right)}\left(K_{\ell,h}^{\left(\boldsymbol{{\tt n}}\right)}\right)^{\top}}{\sqrt{d_{k}}}\bigg)V_{\ell,h}^{\left(\boldsymbol{{\tt n}}\right)}, & \ell=1,3,\\
\text{softmax}\bigg(\frac{Q_{\ell,h}^{\left(\boldsymbol{{\tt n}}\right)}\left(K_{\ell,h}^{\left(\boldsymbol{{\tt n}}\right)}\right)^{\top}}{\sqrt{d_{k}}}+M_{\boldsymbol{{\tt n}}}\bigg)V_{\ell,h}^{\left(\boldsymbol{{\tt n}}\right)}, & \ell=2,
\end{cases}\label{eq: output of multihead, R prime}
\end{alignat}
where $M_{\boldsymbol{{\tt n}}}\in\RR^{d_{k}^{\text{enc}}\times d_{k}^{\text{enc}}}$
denotes the mask, $d_{k}^{\text{enc}}=\frac{3n}{H},$ and $d_{k}^{\text{dec}}=\frac{n}{H}$\footnote{Note, $n$ is divisible by $H$.}.
The query ($Q_{1,h}^{\left(\boldsymbol{{\tt n}}\right)}:\RR^{3n\boldsymbol{\tau}}\to\RR^{d_{k}^{\text{enc}}},\,Q_{2,h}^{\left(\boldsymbol{{\tt n}}\right)},Q_{3,h}^{\left(\boldsymbol{{\tt n}}\right)}:\RR^{n\boldsymbol{\tau}}\to\RR^{d_{k}^{\text{dec}}}$),
key ($K_{1,h}^{\left(\boldsymbol{{\tt n}}\right)}:\RR^{3n\boldsymbol{\tau}}\to\RR^{d_{k}^{\text{enc}}},\,K_{2,h}^{\left(\boldsymbol{{\tt n}}\right)}:\RR^{n\boldsymbol{\tau}}\to\RR^{d_{k}^{\text{dec}}},\,K_{3,h}^{\left(\boldsymbol{{\tt n}}\right)}:\RR^{3n\boldsymbol{\tau}}\to\RR^{d_{k}^{\text{dec}}}$),
and value ($V_{1,h}^{\left(\boldsymbol{{\tt n}}\right)}:\RR^{3n\boldsymbol{\tau}}\to\RR^{d_{k}^{\text{enc}}},\,V_{2,h}^{\left(\boldsymbol{{\tt n}}\right)}:\RR^{n\boldsymbol{\tau}}\to\RR^{d_{k}^{\text{dec}}},\,V_{3,h}^{\left(\boldsymbol{{\tt n}}\right)}:\RR^{3n\boldsymbol{\tau}}\to\RR^{d_{k}^{\text{dec}}}$)
are modeled as

\begin{align}
Q_{\ell,h}^{\left(\boldsymbol{{\tt n}}\right)} & =\left({\tt W}_{\ell,h}^{\left(\boldsymbol{{\tt n}}\right)}\right)^{\top}{\tt C}_{\boldsymbol{{\tt n}},\ell}\left(t\right),\label{eq: query}\\
K_{\ell,h}^{\left(\boldsymbol{{\tt n}}\right)} & =\left(\mathbb{W}_{\ell,h}^{\left(\boldsymbol{{\tt n}}\right)}\right)^{\top}{\tt A}_{\boldsymbol{{\tt n}},\ell}\left(t\right),\label{eq: key}\\
V_{\ell,h}^{\left(\boldsymbol{{\tt n}}\right)} & =\left(\mathbf{W}_{\ell,h}^{\left(\boldsymbol{{\tt n}}\right)}\right)^{\top}{\tt A}_{\boldsymbol{{\tt n}},\ell}\left(t\right),\label{eq: value}
\end{align}
for $\ell\in\left\{ 1,2,3\right\} $, $h\in\left\{ 1,\cdots,H\right\} $,
and $\boldsymbol{{\tt n}}\in\left\{ 1,2,\cdots,N\right\} $. The matrices
${\tt W}_{1,h}^{\left(\boldsymbol{{\tt n}}\right)}\in\RR^{3n\boldsymbol{\tau}\times d_{k}^{\text{enc}}},\,{\tt W}_{2,h}^{\left(\boldsymbol{{\tt n}}\right)},{\tt W}_{3,h}^{\left(\boldsymbol{{\tt n}}\right)}\in\RR^{n\boldsymbol{\tau}\times d_{k}^{\text{dec}}}$
denote the $h^{\text{th}}$ query weights of the multihead of the
$\boldsymbol{{\tt n}}^{\text{th}}$ layer, and these query weights
are vectorized as $\theta_{{\tt W}}\triangleq\bigg[\text{vec}\left({\tt W}_{1,1}^{\left(1\right)}\right)^{\top},\cdots,$
$\text{vec}\left({\tt W}_{1,H}^{\left(1\right)}\right)^{\top},\cdots,$
$\text{vec}\left({\tt W}_{1,1}^{\left(N\right)}\right)^{\top},\cdots\text{,\text{vec}\ensuremath{\left({\tt W}_{1,H}^{\left(N\right)}\right)^{\top}},\ensuremath{\cdots},}$
$\text{vec}\left({\tt W}_{3,1}^{\left(1\right)}\right)^{\top},\cdots,$
$\text{vec}\left({\tt W}_{3,H}^{\left(1\right)}\right)^{\top},\cdots,$
$\text{vec}\left({\tt W}_{3,1}^{\left(N\right)}\right)^{\top},\cdots,$
$\text{vec}\left({\tt W}_{3,H}^{\left(N\right)}\right)^{\top}\bigg]^{\top}$$\in\RR^{11Nn^{2}\boldsymbol{\tau}}$.
The matrices $\mathbb{W}_{1,h}^{\left(\boldsymbol{{\tt n}}\right)}\in\RR^{3n\boldsymbol{\tau}\times d_{k}^{\text{enc}}},\,\mathbb{W}_{2,h}^{\left(\boldsymbol{{\tt n}}\right)}\in\RR^{n\boldsymbol{\tau}\times d_{k}^{\text{dec}}},\,\mathbb{W}_{3,h}^{\left(\boldsymbol{{\tt n}}\right)}\in\RR^{3n\boldsymbol{\tau}\times d_{k}^{\text{dec}}},$
denote the weights associated with the $h^{\text{th}}$ key of the
multihead of the $\boldsymbol{{\tt n}}^{\text{th}}$ layer, and these
key weights are vectorized as $\theta_{\mathbb{W}}\triangleq\bigg[\text{vec}\left(\mathbb{W}_{1,1}^{\left(1\right)}\right)^{\top},\cdots,$
$\text{vec}\left(\mathbb{W}_{1,H}^{\left(1\right)}\right)^{\top},\cdots,$
$\text{vec}\left(\mathbb{W}_{1,1}^{\left(N\right)}\right)^{\top},\cdots,\text{\text{vec}\ensuremath{\left(\mathbb{W}_{1,H}^{\left(N\right)}\right)^{\top}},\ensuremath{\cdots},}$
$\text{vec}\left(\mathbb{W}_{3,1}^{\left(1\right)}\right)^{\top},\cdots,$
$\text{vec}\left(\mathbb{W}_{3,H}^{\left(1\right)}\right)^{\top},\cdots,$
$\text{vec}\left(\mathbb{W}_{3,1}^{\left(N\right)}\right)^{\top},\cdots,$
$\text{vec}\left(\mathbb{W}_{3,H}^{\left(N\right)}\right)^{\top}\bigg]^{\top}$$\in\RR^{13Nn^{2}\boldsymbol{\tau}}$.
The matrices $\mathbf{W}_{1,h}^{\left(\boldsymbol{{\tt n}}\right)}\in\RR^{3n\boldsymbol{\tau}\times d_{k}^{\text{enc}}},\,\mathbf{W}_{2,h}^{\left(\boldsymbol{{\tt n}}\right)}\in\RR^{n\boldsymbol{\tau}\times d_{k}^{\text{dec}}},\,\mathbf{W}_{3,h}^{\left(\boldsymbol{{\tt n}}\right)}\in\RR^{3n\boldsymbol{\tau}\times d_{k}^{\text{dec}}}$
denotes the weights associated with the $h^{\text{th}}$ value of
the multihead of the $\boldsymbol{{\tt n}}^{\text{th}}$ layer, and
these matrices are vectorized as $\theta_{\mathbf{W}}\triangleq\bigg[\text{vec}\left(\mathbf{W}_{1,1}^{\left(1\right)}\right)^{\top},\cdots,$
$\text{vec}\left(\mathbf{W}_{1,H}^{\left(1\right)}\right)^{\top},\cdots,$
$\text{vec}\left(\mathbf{W}_{1,1}^{\left(N\right)}\right)^{\top},\cdots\text{,\text{vec}\ensuremath{\left(\mathbf{W}_{1,H}^{\left(N\right)}\right)^{\top}},\ensuremath{\cdots},}$
$\text{vec}\left(\mathbf{W}_{3,1}^{\left(1\right)}\right)^{\top},\cdots,$
$\text{vec}\left(\mathbf{W}_{3,H}^{\left(1\right)}\right)^{\top},\cdots,$
$\text{vec}\left(\mathbf{W}_{3,1}^{\left(N\right)}\right)^{\top},\cdots,$
$\text{vec}\left(\mathbf{W}_{3,H}^{\left(N\right)}\right)^{\top}\bigg]^{\top}$$\in\RR^{13Nn^{2}\boldsymbol{\tau}}$.
For the Masked Multihead Attention ($\ell=2$), the elements of the
matrix $M_{\boldsymbol{{\tt n}}}$ are calculated as
\begin{equation}
M_{i,j}^{\left(\boldsymbol{{\tt n}}\right)}=\begin{cases}
0, & \text{for}\,i\geq j,\\
-\infty, & \text{for}\,i<j.
\end{cases}\label{eq: mask}
\end{equation}

\subsection{Encoder Block}

In addition to the attention mechanism, each of the $N$ encoder blocks
is composed of two main sub-layers, a multi-head self-attention mechanism,
followed by a fully connected feedforward network. The output of the
multi-head attention sub-layer for the $\boldsymbol{{\tt n}}^{\text{th}}$
encoder block, $R_{\boldsymbol{{\tt n}},\ell}$, serves as the input
to the feedforward sub-layer. The output of each feedforward sub-layer
is denoted as $B_{\boldsymbol{{\tt n}},1}^{\prime}\in\RR^{3n\boldsymbol{\tau}}$
and calculated as

\begin{align}
B_{\boldsymbol{{\tt n}},1}^{\prime} & \triangleq W_{f2}^{\left(\boldsymbol{{\tt n}}\right)}{}^{\top}\sigma_{\text{ReLU}}\left(W_{f1}^{\left(\boldsymbol{{\tt n}}\right)}{}^{\top}R_{\boldsymbol{{\tt n}},1}\left(t\right)\right),\label{eq: Output of encoder feedforward - B1 prime}
\end{align}
where $W_{f1}^{\left(\boldsymbol{{\tt n}}\right)}\in\RR^{3n\boldsymbol{\tau}\times d_{f}}$
and $W_{f2}^{\left(\boldsymbol{{\tt n}}\right)}\in\RR^{d_{f}\times3n\boldsymbol{\tau}}$
denote the feedforward encoder weights, and $d_{f}\in\mathbb{Z}_{>0}$
denotes the size of the hidden layer of the feedforward network. The
encoder feedforward weight matrices are vectorized as {\small{}$\theta_{f1}=\left[\text{vec}\left(W_{f1}^{\left(1\right)}\right)^{\top},\cdots,\text{vec}\left(W_{f1}^{\left(N\right)}\right)^{\top}\right]^{\top}\in\RR^{3nN\boldsymbol{\tau}d_{f}}$
and $\theta_{f2}=\left[\text{vec}\left(W_{f2}^{\left(1\right)}\right)^{\top},\cdots,\text{vec}\left(W_{f2}^{\left(N\right)}\right)^{\top}\right]^{\top}\in\RR^{3nN\boldsymbol{\tau}d_{f}}$.}
The output of the $\boldsymbol{{\tt n}}^{\text{th}}$ encoder block
is denoted as $\Psi^{\left(\boldsymbol{{\tt n}}\right)}\in\RR^{3n\boldsymbol{\tau}}$
and defined as

\begin{align}
\Psi^{\left(\boldsymbol{{\tt n}}\right)} & \left(t\right)=\text{LayerNorm}\left(R_{\boldsymbol{{\tt n}},1}\left(t\right)+B_{\boldsymbol{{\tt n}},1}^{\prime},\gamma_{f}^{\left(\boldsymbol{{\tt n}}\right)},\beta_{f}^{\left(\boldsymbol{{\tt n}}\right)}\right),\label{eq: encder output - Psi}
\end{align}
 where $\gamma_{f}^{\left(\boldsymbol{{\tt n}}\right)},\beta_{f}^{\left(\boldsymbol{{\tt n}}\right)}\in\rr_{>0}$
denote user-selected constants.

\subsection{Decoder Block}

Each of the $N$ identical decoder blocks is composed of three main
sub-layers, a masked multi-head self-attention, a multi-head cross-attention
mechanism that incorporates the $N^{\text{th}}$ encoder's output,
and a fully connected feedforward network. The output of the cross-attention
sub-layer for the $\boldsymbol{{\tt n}}^{\text{th}}$ decoder block,
$R_{\boldsymbol{{\tt n}},3}$, serves as the input to the feedforward
network. The output of the feedforward network is denoted as $B_{\boldsymbol{{\tt n}},2}^{\prime}\in\RR^{n\boldsymbol{\tau}}$
and modeled as
\begin{equation}
B_{\boldsymbol{{\tt n}},2}^{\prime}\triangleq W_{F2}^{\left(\boldsymbol{{\tt n}}\right)}{}^{\top}\sigma_{\text{ReLU}}\left(W_{F1}^{\left(\boldsymbol{{\tt n}}\right)}{}^{\top}R_{\boldsymbol{{\tt n}},3}\left(t\right)\right),\label{eq: Output of decoder feedforward - B2 prime}
\end{equation}
where $W_{F1}^{\left(\boldsymbol{{\tt n}}\right)}\in\RR^{n\boldsymbol{\tau}\times d_{f}}$
and $W_{F2}^{\left(\boldsymbol{{\tt n}}\right)}\in\RR^{d_{f}\times n\boldsymbol{\tau}}$
denote the feedforward decoder weights. The decoder feedforward weights
are defined as {\small{}$\theta_{F1}=\left[\text{vec}\left(W_{F1}^{\left(1\right)}\right)^{\top},\cdots,\text{vec}\left(W_{F1}^{\left(N\right)}\right)^{\top}\right]^{\top}\in\RR^{Nn\boldsymbol{\tau}d_{f}}$
and $\theta_{F2}=\left[\text{vec}\left(W_{F2}^{\left(1\right)}\right)^{\top},\cdots,\text{vec}\left(W_{F2}^{\left(N\right)}\right)^{\top}\right]^{\top}\in\RR^{Nn\boldsymbol{\tau}d_{f}}$.}
The output of the $\boldsymbol{{\tt n}}^{\text{th}}$ encoder block
is denoted as $\Upsilon^{\left(\boldsymbol{{\tt n}}\right)}\in\RR^{n\boldsymbol{\tau}}$
and defined as

\begin{equation}
\Upsilon^{\left(\boldsymbol{{\tt n}}\right)}\left(t\right)=\text{LayerNorm}\left(R_{\boldsymbol{{\tt n}},3}+B_{\boldsymbol{{\tt n}},2}^{\prime},\gamma_{F}^{\left(\boldsymbol{{\tt n}}\right)},\beta_{F}^{\left(\boldsymbol{{\tt n}}\right)}\right),\label{eq: output of decoder - Upsilon}
\end{equation}
where $\gamma_{F}^{\left(\boldsymbol{{\tt n}}\right)},\beta_{F}^{\left(\boldsymbol{{\tt n}}\right)}\in\RR_{>0}$
denote user-selected constants. 

\subsection{LyAT Output\label{subsec:TNN-Output}}

Using the output of the decoder block, the LyAT output is denoted
as $\Phi\left(\boldsymbol{\zeta}_{\text{enc}},\theta\right)\in\RR^{n}$
and defined as
\begin{equation}
\Phi\left(\boldsymbol{\zeta}_{\text{enc}},\theta\right)\tq W_{o}^{\top}\sigma_{\text{ReLU}}\left(\Upsilon^{\left({\tt n}\right)}\right),\label{eq: Lb-TNN output - Phi}
\end{equation}
where $W_{o}\in\RR^{n\boldsymbol{\tau}\times n}$ denotes the output
weight matrix. The stacked matrix of all the weights in the LyAT architecture
is denoted as $\theta\in\RR^{p}$ and defined as
\begin{align*}
\theta & =\left[\theta_{\mathcal{W}}^{\top},\theta_{\mathtt{W}}^{\top},\theta_{\mathbb{W}}^{\top},\theta_{\mathbf{W}}^{\top},\theta_{f1}^{\top},\theta_{f_{2}}^{\top},\theta_{F1}^{\top},\theta_{F2}^{\top},\text{vec}\left(W_{o}\right)^{\top}\right]^{\top},
\end{align*}
where the dimension $p$ is quantified as $p\tq48Nn^{2}\boldsymbol{\tau}+8Nn\boldsymbol{\tau}d_{f}+n^{2}\boldsymbol{\tau}$. 

Recall, the output of the LyAT is used to compensate for the drift
and diffusion uncertainties, $\mathcal{F}:\RR^{3n}\to\RR^{n}$. The
function $\mathcal{F}\left({\tt X}\right)$ can be written as a function
of the encoder input as $\mathcal{F}\left({\tt X}\right)=\mathcal{G}\left(\boldsymbol{\zeta}_{\text{enc}}\right)$
where $\mathcal{G}\left(\boldsymbol{\zeta}_{\text{enc}}\right)$ can
be approximated using the universal function approximation as \cite{Yun.Bhojanapalli.ea2019}
\[
\mathcal{G}\left(\boldsymbol{\zeta}_{\text{enc}}\right)=\Phi\left(\boldsymbol{\zeta}_{\text{enc}},\theta^{*}\right)+\varepsilon\left(\boldsymbol{\zeta}_{\text{enc}}\right).
\]
Since $\mathcal{F}\left({\tt X}\right)=\mathcal{G}\left(\zeta_{\text{enc}}\right)$,
\begin{align}
\mathcal{F}\left({\tt X}\right) & =\Phi\left(\boldsymbol{\zeta}_{\text{enc}},\theta^{*}\right)+\varepsilon\left(\boldsymbol{\zeta}_{\text{enc}}\right),\label{eq: function approximation}
\end{align}
where $\varepsilon\left(\boldsymbol{\zeta}_{\text{enc}}\right):\mathbb{R}^{3n\boldsymbol{\tau}}\rightarrow\mathbb{R}^{n}$
denotes an unknown function representing the reconstruction error
that is bounded by an unknown $\overline{\varepsilon}\in\RR_{>0}$
as
\begin{equation}
\underset{x\in\Omega}{\sup}\left\Vert \varepsilon\left(x\right)\right\Vert \leq\overline{\varepsilon}.\label{eq: varepsilon bound}
\end{equation}

The approximation objective is to determine optimal estimates of $\theta$
such that $\boldsymbol{\zeta}_{\text{enc}}\mapsto\Phi\left(\boldsymbol{\zeta}_{\text{enc}},\hat{\theta}\right)$
approximates ${\tt X}\mapsto\mathcal{F}\left({\tt X}\right)$ with
minimal error for any ${\tt X}\in\Omega$, where $\hat{\theta}$ denote
the adaptive parameter estimates of the ideal weights $\theta^{*}$.
Let $\mathbf{\Omega}\subset\RR^{p}$ denote a user-selected compact
and convex parameter search space with a smooth boundary, satisfying
$\mathbf{0}_{p}\in\text{int}\left(\mathbf{\Omega}\right)$. Additionally,
define $\overline{\theta}\tq\underset{\theta\in\mathbf{\Omega}}{\max}\left\Vert \theta\right\Vert $
to be a bound on the user-selected search space.\footnote{For more information about using a bounded search space for $\hat{\theta}$
and uniqueness, see \cite{Akbari.Patil.ea2025}.}

\subsection{Weight Adaptation Law}

The parameter estimation error is defined as
\begin{equation}
\tilde{\theta}\tq\theta^{*}-\hat{\theta}.\label{eq: parameter estimation error}
\end{equation}
Through the subsequent stochastic Lyapunov stability analysis, the
LyAT parameter update law is designed as
\begin{align}
\dot{\hat{\theta}} & \triangleq\text{proj}\left(\Gamma\Phi^{\prime\top}\left(\boldsymbol{\zeta}_{\text{enc}},\hat{\theta}\right)e-\Gamma\sigma\hat{\theta}\right),\label{eq: Lb-TNN adaptation law}
\end{align}
where $\Gamma\in\rr^{p\times p}$ denotes a user-selected positive-definite
gain matrix, $\sigma\in\RR_{>0}$ denote a user-defined forgetting
factor, proj$\left(\cdot\right)$ denotes a smooth projection operator
defined in \cite[eq. (7)-(11)]{Cai2006a}, which is used to ensure
that $\hat{\theta}$ is bounded as $\left\Vert \hat{\theta}\right\Vert \leq\overline{\theta}$,
and $\Phi^{\prime}\left(\boldsymbol{\zeta}_{\text{enc}},\hat{\theta}\right)\tq\frac{\partial\Phi\left(\boldsymbol{\zeta}_{\text{enc}},\hat{\theta}\right)}{\partial\hat{\theta}}$.

To facilitate the subsequent stability analysis, a first-order Taylor
approximation (cf., \cite{Patil.Le.ea.2022} and \cite{Lewis.Yesildirek.ea1996})
is used to state
\begin{equation}
\Phi\left(\boldsymbol{\zeta}_{\text{enc}},\theta^{*}\right)-\Phi\left(\boldsymbol{\zeta}_{\text{enc}},\hat{\theta}\right)=\Phi^{\prime}\left(\boldsymbol{\zeta}_{\text{enc}},\hat{\theta}\right)\tilde{\theta}+\mathcal{R}\left(\boldsymbol{\zeta}_{\text{enc}},\tilde{\theta}\right),\label{eq: Taylor series approximation}
\end{equation}
where $\mathcal{R}:\RR^{3n\boldsymbol{\tau}}\times\RR^{p}\to\RR^{n}$
denotes Lagrange remainder term. The Lagrange remainder in \eqref{eq: Taylor series approximation}
is unknown; however, the following lemma provides a polynomial bound
for the term.
\begin{lem}
\label{lem: Lagrange remainder bound} There exists a polynomial function
$\rho_{0}:\RR_{\geq0}\to\RR_{\geq0}$ of the form $\rho_{0}\left(\left\Vert x\right\Vert \right)={\tt a}_{2}\left\Vert x\right\Vert ^{2}+{\tt a}_{1}\left\Vert x\right\Vert +{\tt a}_{0}$
with some known constants ${\tt a}_{0},{\tt a}_{1},{\tt a}_{2}\in\RR_{>0}$
such that the Lagrange remainder term can be bounded as $\left\Vert \mathcal{R}\left(x,\tilde{\theta}\right)\right\Vert \leq\rho_{0}\left(\left\Vert x\right\Vert \right)\left\Vert \tilde{\theta}\right\Vert ^{2}$
\cite[Thm. 1]{Patil.Fallin.ea2025}.
\end{lem}

\section{Control Design}

To compensate for the uncertainties that appear in the subsequent
closed-loop error system, the LyAT is incorporated into a controller
designed as
\begin{align}
u & \left(t\right)\triangleq g_{1}^{+}\left(x\right)\left(\dot{x}_{d}-k_{e}e-\Phi\left(\boldsymbol{\zeta}_{\text{enc}},\hat{\theta}\right)\right),\label{eq: Controller}
\end{align}
where $k_{e}\in\mathbb{R}_{>0}$ denotes a user-defined control gain.
Taking the differential of the tracking error in \eqref{eq: Tracking error}
and substituting \eqref{eq: dynamics} and \eqref{eq: Controller},
and canceling the cross terms yields the closed-loop tracking error
\begin{align}
{\rm d}e & =\left(f\left(x\right)-k_{e}e-\Phi\left(\boldsymbol{\zeta}_{\text{enc}},\hat{\theta}\right)\right){\rm d}t+g_{2}\left(x\right)\Sigma\left(t\right){\rm d}\omega.\label{eq:Closed-Loop Error System}
\end{align}

Let $z:\RR_{\geq0}\to\RR^{n+p}$ denote the concatenated error defined
as $z\tq\left[e^{\top},\tilde{\theta}^{\top}\right]^{\top}$. By taking
the differential of the concatenated error $z$ and using \eqref{eq:Closed-Loop Error System},
\eqref{eq: parameter estimation error}, and the chain rule, ${\rm d}z$
is obtained as ${\rm d}z=\left[{\rm d}e^{\top},\dot{\tilde{\theta}}^{\top}{\rm d}t\right]^{\top}$.
Substituting \eqref{eq:Closed-Loop Error System} and \eqref{eq: Lb-TNN adaptation law}
into ${\rm d}z$ yields the closed-loop error system in the form of
a stochastic differential equation as
\begin{equation}
{\rm d}z={\bf F}\left(z\right){\rm d}t+{\bf G}\left(z,t\right){\rm d\omega},\label{eq: System's CLES}
\end{equation}
where ${\bf G}\left(z,t\right)\tq\left[\left(g_{2}\left(x\right)\Sigma\left(t\right)\right)^{\top},\mathbf{0}_{s\times p}^{\top}\right]^{\top}$
and 
\[
{\bf F}\tq\left[\begin{array}{c}
f\left(x\right)-k_{e}e-\Phi\left(\boldsymbol{\zeta}_{\text{enc}},\hat{\theta}\right)\\
-\text{proj}\left(\Gamma\Phi^{\prime\top}\left(\boldsymbol{\zeta}_{\text{enc}},\hat{\theta}\right)e-\Gamma\sigma\hat{\theta}\right)
\end{array}\right].
\]

\section{Stability Analysis\label{sec:Stability-Analysis}}

Consider the Lyapunov function candidate $V_{L}:\mathbb{R}^{n+p}\rightarrow\mathbb{R}_{\geq0}$
defined as

\begin{align}
V_{L}\left(z\right) & \triangleq\frac{1}{2}e^{\top}e+\frac{1}{2}\tilde{\theta}^{\top}\Gamma^{^{-1}}\tilde{\theta}.\label{eq: Lyapunov function}
\end{align}
The Lyapunov function candidate in \eqref{eq: Lyapunov function}
can be bounded as 
\begin{equation}
\alpha_{1}\left\Vert z\right\Vert ^{2}\leq V_{L}\left(z\right)\leq\alpha_{2}\left\Vert z\right\Vert ^{2},\label{eq: Raleigh-Reitx}
\end{equation}
where $\alpha_{1}\tq\frac{1}{2}\min\left(1,\Gamma^{-1}\right)$ and
$\alpha_{2}\tq\max\left(1,\Gamma^{-1}\right)$. 

For the universal function approximation property to hold, it is necessary
to ensure $\boldsymbol{\zeta}_{\text{enc}}\in\Omega$, for all $t\in\RR_{\geq0}$.
To guarantee that $\boldsymbol{\zeta}_{\text{enc}}\in\Omega$, the
subsequent stability analysis contains $z$ in a compact domain, for
all $t\in\RR_{\geq0}$. Consider the compact domain $\mathcal{D}\tq\left\{ \iota\in\RR^{n+p}:\left\Vert \iota\right\Vert \leq\rho^{-1}\left({\tt b}_{0}-{\tt b}_{2}\right)\right\} $
in which $z$ is supposed to lie for all time $t\in\RR_{\geq0}$,
where ${\tt b}_{0}\tq\min\left(\frac{k_{e}}{2},\frac{\sigma}{2}\right)$,
${\tt b}_{2}\in\RR_{>0}$ denotes the desired rate of convergence,
and $\rho\left(\left\Vert z\right\Vert \right)\tq\rho_{1}\left(\left\Vert z\right\Vert \right)\left\Vert z\right\Vert $
is invertible. It follows that if $\left\Vert z\right\Vert \in\mathcal{D}$,
then using \eqref{eq: Tracking error}, triangle inequality, and the
bound on the desired trajectory yields
\begin{align}
\left\Vert {\tt X}_{i}\right\Vert  & \le\left\Vert x_{i}\right\Vert +\left\Vert x_{d,i}\right\Vert +\left\Vert e_{i}\right\Vert \leq2\left\Vert e_{i}\right\Vert +2\left\Vert x_{d,i}\right\Vert \nonumber \\
 & \leq2\rho^{-1}\left({\tt b}_{0}-{\tt b}_{2}\right)+2\overline{x_{d}},\label{eq: bound on X_i}
\end{align}
for all $i\in\left\{ 1,2,\cdots,\boldsymbol{\tau}\right\} $. Since
\eqref{eq: bound on X_i} holds for all $i\in\left\{ 1,2,\cdots,\boldsymbol{\tau}\right\} $,
then $\left\Vert \boldsymbol{\zeta}_{\text{enc}}\right\Vert \leq\left\Vert {\tt X}_{1}\right\Vert +\cdots\left\Vert {\tt X}\right\Vert \leq2\boldsymbol{\tau}\rho^{-1}\left(\kappa_{0}-\kappa_{2}\right)+2\boldsymbol{\tau}\overline{x_{d}}$.
Therefore, select $\Omega\tq\left\{ \iota\in\RR^{3n\boldsymbol{\tau}}:\left\Vert \iota\right\Vert \leq2\boldsymbol{\tau}\rho^{-1}\left(\kappa_{0}-\kappa_{2}\right)+2\boldsymbol{\tau}\overline{x_{d}}\right\} $.
Then, $z\in\mathcal{D}$ implies $\boldsymbol{\zeta}_{\text{enc}}\in\Omega$.
In the subsequent analysis, it is shown that if $z\left(0\right)\in\mathcal{S}$,
then $z\left(t\right)\in\mathcal{D}$, for all $t\in\RR_{\geq0}$,
where $\mathcal{S}\subset\mathcal{D}$ is defined
\begin{equation}
\mathcal{S}\tq\left\{ \iota\in\RR^{n+p}:\left\Vert \iota\right\Vert \leq\sqrt{\frac{\alpha_{1}}{\alpha_{2}}\left(\rho^{-1}\left({\tt b}_{0}-{\tt b}_{2}\right)\right)^{2}-\frac{{\tt b}_{1}}{{\tt b}_{2}}}\right\} ,\label{eq:Chi Definition}
\end{equation}
where ${\tt b}_{1}\tq\frac{\overline{\varepsilon}^{2}}{2k_{e}}+\frac{\sigma}{2}\overline{\theta}^{2}+\frac{1}{2}\left\Vert \Sigma\right\Vert _{F\infty}^{2}\overline{g}^{2}$.
Note that the set $\mathcal{S}$ exists when ${\tt b}_{0}\geq{\tt b}_{2}+\rho\left(\sqrt{\frac{\alpha_{2}}{\alpha_{1}}\frac{{\tt b}_{1}}{{\tt b}_{2}}}\right)$.
Additionally, $\mathcal{S}$ can be made arbitrarily large to include
any initial condition $z\left(0\right)$ such that ${\tt b}_{0}$
satisfies the gain condition
\begin{equation}
{\tt b}_{0}\geq\rho\left(\sqrt{\frac{\alpha_{2}}{\alpha_{1}}\left\Vert z\right\Vert ^{2}+\frac{\alpha_{2}}{\alpha_{1}}\frac{{\tt b}_{1}}{{\tt b}_{2}}}\right)+{\tt b}_{2}.\label{eq: gain condition}
\end{equation}
The gain condition in \eqref{eq: gain condition} is equivalent to
stating the set $\mathcal{S}$ is not a null set and contains the
initial condition $z\left(0\right)$.

Since $\mathcal{S}\subset\mathcal{D}$, $z\left(0\right)\in\mathcal{S}$
implies $z\left(0\right)\in\mathring{\mathcal{D}}$ where $\mathring{\mathcal{D}}$
denotes the interior of $\mathcal{D}$. The solution $t\mapsto z\left(t\right)$
is assumed to be continuous on a time-interval $\mathcal{T}\triangleq\left[0,t_{1}\right]$
with probability one\footnote{This is a standard assumption in the analysis of stochastic systems
(cf. \cite[Page 36, Ass. (A2)]{Kushner1971}).} such that $z\left(t\right)\in\mathcal{D}$ for all $t\in\mathcal{T}$.
It follows that $x\left(t\right)\in\Omega$, for all $t\in\mathcal{T}$,
therefore the universal function approximation property holds over
this time interval. In the subsequent stability analysis, the probabilistic
convergence properties of the solutions are analyzed. Using Lemma
\ref{thm:probability} and to facilitate the analysis, the risk of
$z$ to escape $\mathcal{D}$ is denoted as $\vartheta$, which is
defined as
\begin{align}
\vartheta & \tq\frac{1}{\alpha_{1}\left(\rho^{-1}\left({\tt b}_{0}-{\tt b}_{2}\right)\right)^{2}}V_{L}\left(z\left(0\right)\right)\nonumber \\
 & \quad+\frac{1}{\lambda}V_{L}\left(z\left(0\right)\right)\exp\left(-\frac{{\tt b}_{2}}{\alpha_{2}}t\right)+\frac{\alpha_{2}{\tt b}_{1}}{\lambda{\tt b}_{2}},\label{eq: Escape risk}
\end{align}
 where $\lambda\in\left[\frac{\alpha_{2}{\tt b}_{1}}{{\tt b}_{2}},\alpha_{1}\left(\rho^{-1}\left({\tt b}_{0}-{\tt b}_{2}\right)\right)^{2}\right].$
\begin{thm}
\label{thm: main stability theorem}Consider the dynamical system
in \eqref{eq: dynamics}. Given the gain condition in \eqref{eq: gain condition},
for any initial conditions of the process $z\left(0\right)\in\mathcal{S}$,
the LyAT parameter adaptation law and the controller respectively
given by \eqref{eq: Lb-TNN adaptation law} and \eqref{eq: Controller}
guarantee that the Ito process $z\left(t\right)$ is uniformly ultimately
bounded in probability (UUB-p, c.f., \cite[Def. 1]{Akbari.Nino.ea2024})
in the sense that
\begin{equation}
{\rm P}\left(\underset{t\leq s<\infty}{\sup}\left\Vert z\left(s\right)\right\Vert \leq\sqrt{\frac{\lambda}{\alpha_{1}}}\right)\geq1-\vartheta.\label{eq: Stability result}
\end{equation}
\end{thm}
\begin{IEEEproof}
Taking the infinitesimal generator of the candidate Lyapunov function
in \eqref{eq: Lyapunov function} yields
\begin{equation}
\mathcal{L}V_{L}\left(z\right)=\frac{\partial V_{L}}{\partial z}{\bf F}\left(z\right)+\frac{1}{2}\text{tr}\left({\bf G}^{\top}\left(z,t\right)\frac{\partial^{2}V_{L}}{\partial z^{2}}{\bf G}\left(z,t\right)\right).\label{eq: Ito's derivation of V_L}
\end{equation}
Substituting ${\bf F}$, ${\bf G}$, $\frac{\partial V_{L}}{\partial z}$,
and $\frac{\partial^{2}V_{L}}{\partial z^{2}}$ into \eqref{eq: Ito's derivation of V_L}
yields
\begin{align}
\mathcal{L}V_{L}\left(z\right) & =e^{\top}\left(f\left(x\right)-k_{e}e-\Phi\left(\boldsymbol{\zeta}_{\text{enc}},\hat{\theta}\right)\right)\nonumber \\
 & \hspace{1em}-\tilde{\theta}^{\top}\Gamma^{-1}\text{proj}\left(\Gamma\Phi^{\prime\top}\left(\boldsymbol{\zeta}_{\text{enc}},\hat{\theta}\right)e-\Gamma\sigma\hat{\theta}\right)\nonumber \\
 & \hspace{1em}+\frac{1}{2}\text{tr}\left(\Sigma^{\top}\left(t\right)g_{2}^{\top}\left(x\right)g_{2}\left(x\right)\Sigma\left(t\right)\right).\label{eq: Substituting CLES into LV}
\end{align}
From \cite[P2 in Thm. 1]{Cai2006a}, $\left(\cdot\right)\leq\text{proj}\left(\cdot\right)$.
Using this lower bound of proj$\left(\cdot\right)$ and applying the
trace property from \eqref{eq: order of multiplication property}
on \eqref{eq: Substituting CLES into LV} yields 
\begin{align}
\mathcal{L}V_{L} & \leq e^{\top}\left(f\left(x\right)-k_{e}e-\Phi\left(\boldsymbol{\zeta}_{\text{enc}},\hat{\theta}\right)\right)-\tilde{\theta}^{\top}\Phi^{\prime\top}\left(\boldsymbol{\zeta}_{\text{enc}},\hat{\theta}\right)e\nonumber \\
 & \hspace{1em}+\sigma\tilde{\theta}^{\top}\hat{\theta}+\frac{1}{2}\text{tr}\left(g_{2}^{\top}\left(x\right)g_{2}\left(x\right)\Sigma\left(t\right)\Sigma^{\top}\left(t\right)\right).\label{eq: upperbounding proj}
\end{align}
Using the definition of the Frobenius norm on the term $\text{tr}\left(g_{2}^{\top}\left(x\right)g_{2}\left(x\right)\Sigma\left(t\right)\Sigma^{\top}\left(t\right)\right)$
yields 
\begin{align}
 & \text{tr}\left(g_{2}^{\top}\left(x\right)g_{2}\left(x\right)\Sigma\left(t\right)\Sigma^{\top}\left(t\right)\right)\nonumber \\
 & \hspace{1em}=\text{tr}\left(\Sigma^{\top}\left(t\right)g_{2}^{\top}\left(x\right)g_{2}\left(x\right)\Sigma\left(t\right)\right)\nonumber \\
 & \hspace{1em}=\text{tr}\left(\left(g_{2}\left(x\right)\Sigma\left(t\right)\right)^{\top}g_{2}\left(x\right)\Sigma\left(t\right)\right)\nonumber \\
 & \hspace{1em}=\left\Vert g_{2}\left(x\right)\Sigma\left(t\right)\right\Vert _{F}^{2}.\label{eq: ggss}
\end{align}
Applying the Cauchy-Schwarz inequality \cite[Page 189]{Axler2024}
to \eqref{eq: ggss} yields
\begin{align}
\left\Vert g_{2}\left(x\right)\Sigma\left(t\right)\right\Vert _{F}^{2} & \leq\left\Vert g_{2}\left(x\right)\right\Vert _{F}^{2}\left\Vert \Sigma\left(t\right)\right\Vert _{F}^{2}\nonumber \\
 & =\text{tr}\left(g_{2}^{\top}\left(x\right)g_{2}\left(x\right)\right)\left\Vert \Sigma\left(t\right)\right\Vert _{F}^{2}.\label{eq: holder on ggss}
\end{align}
Using \eqref{eq: separating trace inequality}, \eqref{eq: ggss},
\eqref{eq: holder on ggss}, and the fact that $\left\Vert \Sigma\left(t\right)\right\Vert _{F}^{2}\leq\underset{t\in\RR_{\geq0}}{\sup}\left\Vert \Sigma\left(t\right)\right\Vert _{F}^{2}\triangleq\left\Vert \Sigma\right\Vert _{F\infty}^{2}$,
the term $\text{tr}\left(g_{2}^{\top}\left(x\right)g_{2}\left(x\right)\Sigma\left(t\right)\Sigma^{\top}\left(t\right)\right)$
in \eqref{eq: upperbounding proj} is upper bounded as 
\begin{equation}
\text{tr}\left(g_{2}^{\top}\left(x\right)g_{2}\left(x\right)\Sigma\left(t\right)\Sigma^{\top}\left(t\right)\right)\leq\text{tr}\left(g_{2}^{\top}\left(x\right)g_{2}\left(x\right)\right)\left\Vert \Sigma\right\Vert _{F\infty}^{2}.\label{eq: ggss final}
\end{equation}
Thus, applying \eqref{eq: ggss final} to \eqref{eq: upperbounding proj},
applying the trace-to-vector property in \eqref{eq: trace to vec}
to $\text{tr}\left(g_{2}^{\top}g_{2}\right)$, and substituting \eqref{eq: vectorized g2}
into \eqref{eq: upperbounding proj} yields 
\begin{align}
\mathcal{L}V_{L} & \leq e^{\top}\left(f\left(x\right)-k_{e}e-\Phi\left(\boldsymbol{\zeta}_{\text{enc}},\hat{\theta}\right)\right)-\tilde{\theta}^{\top}\Phi^{\prime\top}\left(\boldsymbol{\zeta}_{\text{enc}},\hat{\theta}\right)e\nonumber \\
 & \hspace{1em}+\sigma\tilde{\theta}^{\top}\hat{\theta}+\frac{1}{2}\left\Vert \Sigma\right\Vert _{F\infty}^{2}\bigg(e^{\top}G_{2}^{\top}\left(e,x_{d}\right)G_{2}\left(e,x_{d}\right)e\nonumber \\
 & \hspace{1em}+2e^{\top}G_{2}^{\top}\left(e,x_{d}\right)\text{vec}\left(g_{2}\left(x_{d}\right)\right)\nonumber \\
 & \hspace{1em}+\text{vec}\left(g_{2}\left(x_{d}\right)\right)^{\top}\text{vec}\left(g_{2}\left(x_{d}\right)\right)\bigg).\label{eq: sub. taylor model of g2}
\end{align}
Since $e^{\top}G_{2}^{\top}(e,x_{d})G_{2}(e,x_{d})e=\text{tr}\{e^{\top}G_{2}^{\top}(e,x_{d})G_{2}(e,x_{d})e\}$,
using \eqref{eq: order of multiplication property} and \eqref{eq: separating trace inequality}
and upper-bounding $\tvec\left(g_{2}\left(x_{d}\right)\right)$ as
$\tvec\left(g_{2}\left(x_{d}\right)\right)\leq\left\Vert \tvec\left(g_{2}\left(x_{d}\right)\right)\right\Vert \leq\overline{g}$,
yields
\begin{align}
\mathcal{L}V_{L} & \leq e^{\top}\left(f\left(x\right)-k_{e}e-\Phi\left(\boldsymbol{\zeta}_{\text{enc}},\hat{\theta}\right)\right)-\tilde{\theta}^{\top}\Phi^{\prime\top}\left(\boldsymbol{\zeta}_{\text{enc}},\hat{\theta}\right)e\nonumber \\
 & +\sigma\tilde{\theta}^{\top}\hat{\theta}+\frac{1}{2}\left\Vert \Sigma\right\Vert _{F\infty}^{2}e^{\top}e\text{tr}\left\{ G_{2}^{\top}\left(e,x_{d}\right)G_{2}\left(e,x_{d}\right)\right\} \nonumber \\
 & +\left\Vert \Sigma\right\Vert _{F\infty}^{2}e^{\top}G_{2}^{\top}\left(e,x_{d}\right)\text{vec}\left(g_{2}\left(x_{d}\right)\right)+\frac{1}{2}\left\Vert \Sigma\right\Vert _{F\infty}^{2}\overline{g}^{2}.\label{eq: getting LV to the F form}
\end{align}
Applying the definition of $\mathcal{F}$ and then using \eqref{eq: function approximation}
yields
\begin{align}
\mathcal{L}V_{L} & \leq e^{\top}\left(-k_{e}e+\Phi\left(\boldsymbol{\zeta}_{\text{enc}},\theta^{*}\right)-\Phi\left(\boldsymbol{\zeta}_{\text{enc}},\hat{\theta}\right)+\varepsilon\left(\boldsymbol{\zeta}_{\text{enc}}\right)\right)\nonumber \\
 & \hspace{1em}-\tilde{\theta}^{\top}\Phi^{\prime\top}\left(\boldsymbol{\zeta}_{\text{enc}},\hat{\theta}\right)e+\sigma\tilde{\theta}^{\top}\hat{\theta}+\frac{1}{2}\left\Vert \Sigma\right\Vert _{F\infty}^{2}\overline{g}^{2}.\label{eq: applying TNN}
\end{align}
Applying the Taylor series approximation in \eqref{eq: Taylor series approximation}
on \eqref{eq: applying TNN} and canceling cross terms yields
\begin{equation}
\mathcal{L}V_{L}\leq e^{\top}\left(-k_{e}e+\Delta\left(\boldsymbol{\zeta}_{\text{enc}},\tilde{\theta}\right)\right)+\sigma\tilde{\theta}^{\top}\hat{\theta}+\frac{1}{2}\left\Vert \Sigma\right\Vert _{F\infty}^{2}\overline{g}^{2},\label{eq: LV after taylor}
\end{equation}
where $\Delta:\RR^{3n\boldsymbol{\tau}}\times\RR^{p}\to\RR^{n}$ is
defined as $\text{\ensuremath{\Delta\left(\boldsymbol{\zeta}_{\text{enc}},\tilde{\theta}\right)}}\tq\varepsilon\left(\boldsymbol{\zeta}_{\text{enc}}\right)+\mathcal{R}\left(\boldsymbol{\zeta}_{\text{enc}},\tilde{\theta}\right)$.
By using \eqref{eq: Tracking error}, \eqref{eq: varepsilon bound},
Lemma \ref{lem: Lagrange remainder bound}, the triangle inequality,
and boundedness of desired trajectory, the term $\Delta\left(\boldsymbol{\zeta}_{\text{enc}},\tilde{\theta}\right)$
can be bounded as 
\begin{equation}
\left\Vert \Delta\left(\boldsymbol{\zeta}_{\text{enc}},\tilde{\theta}\right)\right\Vert \leq\overline{\varepsilon}+\rho_{0}\left(2\sum_{i=1}^{\boldsymbol{\tau}}\left\Vert e_{i}\right\Vert +2\boldsymbol{\tau}\overline{x_{d}}\right)\left\Vert \tilde{\theta}\right\Vert ^{2},\label{eq: Delta upperbound}
\end{equation}
for all $t\in\mathcal{T}$. Since $\rho_{0}$ is strictly increasing,
the term $\rho_{0}\left(2\sum_{i=1}^{\boldsymbol{\tau}}\left\Vert e_{i}\right\Vert +2\boldsymbol{\tau}\overline{x_{d}}\right)$
can be upper-bounded as $\rho_{0}\left(2\sum_{i=1}^{\boldsymbol{\tau}}\left\Vert e_{i}\right\Vert +2\boldsymbol{\tau}\overline{x_{d}}\right)\leq\rho_{0}\left(2\sum_{i=1}^{\boldsymbol{\tau}}\left\Vert z_{i}\right\Vert +2\boldsymbol{\tau}\overline{x_{d}}\right)$.
Using this upper bound and concatenated error definition, \eqref{eq: Delta upperbound}
is further upper-bounded as
\begin{equation}
\left\Vert \Delta\left(\boldsymbol{\zeta}_{\text{enc}},\tilde{\theta}\right)\right\Vert \leq\overline{\varepsilon}+\rho_{1}\left(\left\Vert z_{i}\right\Vert \right)\left\Vert z\right\Vert ^{2},\label{eq: Final Delta upperbound}
\end{equation}
where $\rho_{1}\left(\left\Vert z_{i}\right\Vert \right)=\rho_{0}\left(2\sum_{i=1}^{\boldsymbol{\tau}}\left\Vert z_{i}\right\Vert +2\boldsymbol{\tau}\overline{x_{d}}\right)$,
for all $t\in\mathcal{T}$. Applying \eqref{eq: Final Delta upperbound}
and \eqref{eq: parameter estimation error} to \eqref{eq: LV after taylor}
yields
\begin{align}
\mathcal{L}V_{L} & \leq-k_{e}\left\Vert e\right\Vert ^{2}+\left|e^{\top}\right|\overline{\varepsilon}+\left|e^{\top}\right|\rho_{1}\left(\left\Vert z_{i}\right\Vert \right)\left\Vert z\right\Vert ^{2}\nonumber \\
 & -\sigma\left\Vert \tilde{\theta}\right\Vert ^{2}+\sigma\left\Vert \tilde{\theta}\right\Vert \left\Vert \theta^{*}\right\Vert +\frac{1}{2}\left\Vert \Sigma\right\Vert _{F\infty}^{2}\overline{g}^{2},\label{eq: LV breaking sigma mod}
\end{align}
for all $t\in\mathcal{T}$. Let $\rho\left(\left\Vert z_{i}\right\Vert \right)\triangleq\rho_{1}\left(\left\Vert z_{i}\right\Vert \right)\left\Vert z\right\Vert $.
From \eqref{eq: LV breaking sigma mod}, using the bound of $\theta^{*}$
and applying Young's inequality on $\overline{\varepsilon}\left\Vert e\right\Vert $
and $\sigma\overline{\theta}\left\Vert \tilde{\theta}\right\Vert $
yields
\begin{equation}
\mathcal{L}V_{L}\leq-\left({\tt b_{0}-}\rho\left(\left\Vert z_{i}\right\Vert \right)\right)\left\Vert z\right\Vert ^{2}+{\tt b}_{1},\label{eq: LV almost final}
\end{equation}
for all $t\in\mathcal{T}$. Based on the definition of $\mathcal{D}$,
it follows that for each $i\in\left\{ 1,\cdots,\boldsymbol{\tau}\right\} $
that $\rho\left(\left\Vert z_{i}\right\Vert \right)\leq{\tt b}_{0}-{\tt b}_{2}$.
Therefore, 
\begin{equation}
\mathcal{L}V_{L}\leq-{\tt b}_{2}\left\Vert z\right\Vert ^{2}+{\tt b}_{1},\label{eq: LV_L are we there yet}
\end{equation}
for all $t\in\mathcal{T}$. Applying \eqref{eq: Raleigh-Reitx} to
\eqref{eq: LV_L are we there yet} yields
\begin{equation}
\mathcal{L}V_{L}\leq-\frac{{\tt b}_{2}}{\alpha_{2}}V_{L}+{\tt b}_{1},\label{eq: LV_L FINALLL!!-1}
\end{equation}
for all $t\in\mathcal{T}$.

Since $V_{L}\left(0\right)=0$, $V_{L}\in\mathcal{C}^{2}$, and $z$
is a continuous strong Markov process, based on \eqref{eq: LV_L FINALLL!!-1},
then \cite[Lemma 1]{Akbari.Nino.ea2024} can be invoked to state 
\[
{\rm P}\left(\underset{t\leq s\leq\infty}{\sup}V_{L}\left(z\left(s\right)\right)\geq\lambda\right)\leq\vartheta,
\]
which is equivalent to 
\begin{equation}
{\rm P}\left(\underset{t\leq s\leq\infty}{\sup}V_{L}\left(z\left(s\right)\right)<\lambda\right)\geq1-\vartheta,\label{eq: probability of VL being bounded}
\end{equation}
for all $t\in\mathcal{T}$. From \eqref{eq: Raleigh-Reitx}, ${\rm P}\left(\underset{t\leq s<\infty}{\sup}\left\Vert z\left(s\right)\right\Vert ^{2}<\frac{\lambda}{\alpha_{1}}\right)\geq{\rm P}\left(\underset{t\leq s<\infty}{\sup}V_{L}\left(z\left(s\right)\right)<\lambda\right)$.
Therefore, using \eqref{eq: probability of VL being bounded} and
Lemma \ref{thm:probability} yields \eqref{eq: Stability result},
for all $t\in\mathcal{T}$. From \eqref{eq: Stability result} and
\cite[Def. 1]{Akbari.Nino.ea2024}, the solution $z\left(t\right)$
is UUB-p, for all $t\in\mathcal{T}$.
\end{IEEEproof}
\begin{rem}
The tunable paramters such as $k_{e}$ directly influences the parameters
${\tt b}_{0}$, ${\tt b}_{1}$, and ${\tt b}_{2}$, which in turn
affect $\rho^{-1}$ and the size of the initialization set. By making
$\rho^{-1}$ larger, the admissible range for $\lambda$ can b increased.
From \eqref{eq: Escape risk}, larger values of $\lambda$ and $\rho^{-1}$
result in smaller $\vartheta$, thereby increasing the probability
$1-\vartheta$ that $z$ remains bounded. 
\end{rem}
Let ${\tt S}_{1}\tq\left\{ z:\left\Vert z\left(t\right)\right\Vert <\sqrt{\frac{\lambda}{\alpha_{1}}}\right\} $
and ${\tt S}_{2}\tq\left\{ z:\left\Vert e\left(t\right)\right\Vert <\sqrt{\frac{\lambda}{\alpha_{1}}}\right\} $.
Since $S_{1}\subseteq S_{2}$, the monotonicity p.roperty in \eqref{eq:monotonicity}
yields ${\rm P}\left(\underset{t\leq s<\infty}{\sup}\left\Vert e\left(s\right)\right\Vert \leq\sqrt{\frac{\lambda}{\alpha_{1}}}\right)\geq{\rm P}\left(\underset{t\leq s<\infty}{\sup}\left\Vert z\left(s\right)\right\Vert \leq\sqrt{\frac{\lambda}{\alpha_{1}}}\right)\geq1-\vartheta$,
for all $i\in\left\{ 1,2,\cdots,\boldsymbol{\tau}\right\} $. Let
${\tt S}_{3,i}\tq\left\{ z:\left\Vert z_{i}\right\Vert <\sqrt{\frac{\lambda}{\alpha_{1}}}\right\} $,
for all $i\in\left\{ 1,2,\cdots,\boldsymbol{\tau}\right\} $. Since
${\tt S}_{1}\subseteq{\tt S}_{3,i}$, the monotonicity property in
\eqref{eq:monotonicity} yields ${\rm P}\left(\underset{t\leq s<\infty}{\sup}\left\Vert z_{i}\right\Vert \leq\sqrt{\frac{\lambda}{\alpha_{1}}}\right)\geq{\rm P}\left(\underset{t\leq s<\infty}{\sup}\left\Vert z\left(s\right)\right\Vert \leq\sqrt{\frac{\lambda}{\alpha_{1}}}\right)\geq1-\vartheta$,
for all $i\in\left\{ 1,2,\cdots,\boldsymbol{\tau}\right\} $. Let
${\tt S}_{4,i}\tq\left\{ z:\left\Vert e_{i}\right\Vert <\sqrt{\frac{\lambda}{\alpha_{1}}}\right\} $,
for all $i\in\left\{ 1,2,\cdots,\boldsymbol{\tau}\right\} $. Since
${\tt S}_{3,i}\subseteq{\tt S}_{4,i}$, the monotonicity property
in \eqref{eq:monotonicity} yields ${\rm P}\left(\underset{t\leq s<\infty}{\sup}\left\Vert e_{i}\right\Vert <\sqrt{\frac{\lambda}{\alpha_{1}}}\right)\geq{\rm P}\left(\underset{t\leq s<\infty}{\sup}\left\Vert z_{i}\right\Vert <\sqrt{\frac{\lambda}{\alpha_{1}}}\right)\geq1-\vartheta$,
for all $i\in\left\{ 1,2,\cdots,\boldsymbol{\tau}\right\} $. Let
${\tt S}_{5,i}\tq\left\{ z:\left\Vert x_{i}\right\Vert <\sqrt{\frac{\lambda}{\alpha_{1}}}+\overline{x_{d}}\right\} $,
for all $i\in\left\{ 1,2,\cdots,\boldsymbol{\tau}\right\} $. Since
${\tt S}_{4,i}\subseteq{\tt S}_{5,i}$, the monotonicity property
in \eqref{eq:monotonicity} yields ${\rm P}\left(\underset{t\leq s<\infty}{\sup}\left\Vert x_{i}\right\Vert <\sqrt{\frac{\lambda}{\alpha_{1}}}+\overline{x_{d}}\right)\geq{\rm P}\left(\underset{t\leq s<\infty}{\sup}\left\Vert e_{i}\right\Vert <\sqrt{\frac{\lambda}{\alpha_{1}}}\right)\geq1-\vartheta$,
for all $i\in\left\{ 1,2,\cdots,\boldsymbol{\tau}\right\} $. Since
$x_{1},\cdots,x_{\boldsymbol{\tau}}$ are not independent of each
other and ${\rm P}\left(\underset{t\leq s<\infty}{\sup}\left\Vert x_{i}\right\Vert \leq\sqrt{\frac{\lambda}{\alpha_{1}}}\right)\geq1-\vartheta$,
it can be stated that 
\begin{gather}
{\rm P}\big(\big\{\underset{t\leq s<\infty}{\sup}\left\Vert x_{1}\right\Vert \leq\sqrt{\frac{\lambda}{\alpha_{1}}}\big\}\cap\cdots\cap\big\{\underset{t\leq s<\infty}{\sup}\left\Vert x_{\boldsymbol{\tau}}\right\Vert \leq\sqrt{\frac{\lambda}{\alpha_{1}}\big\}}\big)\nonumber \\
\geq\underset{i\in\left\{ 1,\cdots,\boldsymbol{\tau}\right\} }{\min}\bigg({\rm P}\big(\underset{t\leq s<\infty}{\sup}\left\Vert x_{1}\right\Vert \leq\sqrt{\frac{\lambda}{\alpha_{1}}}\big),\cdots,\nonumber \\
{\rm P}\big(\underset{t\leq s<\infty}{\sup}\left\Vert x_{\boldsymbol{\tau}}\right\Vert \leq\sqrt{\frac{\lambda}{\alpha_{1}}}\big)\bigg)\geq1-\vartheta.\label{eq: probabilistic bound on window x}
\end{gather}
Since $e_{1},\cdots,e_{\boldsymbol{\tau}}$ are not independent of
each other and ${\rm P}\left(\underset{t\leq s<\infty}{\sup}\left\Vert e_{i}\right\Vert \leq\sqrt{\frac{\lambda}{\alpha_{1}}}\right)\geq1-\vartheta$,
it can be stated that 
\begin{gather}
{\rm P}\big(\big\{\underset{t\leq s<\infty}{\sup}\left\Vert e_{1}\right\Vert \leq\sqrt{\frac{\lambda}{\alpha_{1}}}\big\}\cap\cdots\cap\big\{\underset{t\leq s<\infty}{\sup}\left\Vert e_{\boldsymbol{\tau}}\right\Vert \leq\sqrt{\frac{\lambda}{\alpha_{1}}}\big\}\big)\nonumber \\
\geq\underset{i\in\left\{ 1,\cdots,\boldsymbol{\tau}\right\} }{\min}\bigg({\rm P}\big(\underset{t\leq s<\infty}{\sup}\left\Vert e_{1}\right\Vert \leq\sqrt{\frac{\lambda}{\alpha_{1}}}\big),\cdots,\nonumber \\
{\rm P}\big(\underset{t\leq s<\infty}{\sup}\left\Vert e_{\boldsymbol{\tau}}\right\Vert \leq\sqrt{\frac{\lambda}{\alpha_{1}}}\big)\bigg)\geq1-\vartheta.\label{eq: probabilistic bound on window e}
\end{gather}
From the boundedness of $x_{d}$ and using \eqref{eq: probabilistic bound on window x},
\eqref{eq: probabilistic bound on window e}, there exists a constant
$\overline{\boldsymbol{\zeta}_{\text{enc}}}$ such that ${\rm P}\left(\underset{t\leq s<\infty}{\sup}\left\Vert \boldsymbol{\zeta}_{\text{enc}}\right\Vert <\overline{\boldsymbol{\zeta}_{\text{enc}}}\right)\geq1-\vartheta$.
Since ${\rm P}\left(\underset{t\leq s<\infty}{\sup}\left\Vert \boldsymbol{\zeta}_{\text{enc}}\right\Vert <\overline{\boldsymbol{\zeta}_{\text{enc}}}\right)\geq1-\vartheta$
and $\left\Vert \hat{\theta}\right\Vert \leq\overline{\theta}$, and
based on the smoothness of $\Phi\left(\boldsymbol{\zeta}_{\text{enc}},\hat{\theta}\right)$,
there exists a constant $\overline{\Phi}\in\RR_{>0}$ such that ${\rm P}\left(\underset{t\leq s<\infty}{\sup}\left\Vert \Phi\left(\boldsymbol{\zeta}_{\text{enc}},\hat{\theta}\left(s\right)\right)\right\Vert \leq\overline{\Phi}\right)\geq1-\vartheta$.
Since ${\rm P}\left(\underset{t\leq s<\infty}{\sup}\left\Vert \Phi\left(\boldsymbol{\zeta}_{\text{enc}},\hat{\theta}\left(s\right)\right)\right\Vert \leq\overline{\Phi}\right)\geq1-\vartheta$,
${\rm P}\left(\underset{t\leq s<\infty}{\sup}\left\Vert e\left(s\right)\right\Vert <\sqrt{\frac{\lambda}{\alpha_{1}}}\right)\geq1-\vartheta$,
and $\left\Vert x_{d}\right\Vert \leq\overline{x_{d}}$, using \eqref{eq: Controller}
yields ${\rm P}\left(\underset{t\leq s<\infty}{\sup}\left\Vert u\left(s\right)\right\Vert \leq\overline{u}\right)\geq1-\vartheta$,
for some constant $\overline{u}\in\RR_{>0}$. Therefore, all implemented
signals are bounded with probability of $1-\vartheta$.

\section{Experiment}

\subsection{Experimental Testbed}

To verify the efficacy of the proposed LyAT control method, a 4-minute
experimental validation is performed on a Freefly Astro quadrotor
at the University of Florida's Autonomy Park outdoor facility. The
Freefly Astro is a professional-grade quadcopter with an unfolded
diameter of 930 mm (1407 mm including propellers), powered by four
Freefly 7010 motors with 21\texttimes 7 inches carbon fiber reinforced
nylon folding propellers. The aircraft has a maximum gross takeoff
weight of 8700 g, and the flight controller is a Freefly custom-designed
Skynode running Auterion Enterprise PX4 firmware.

\begin{figure}[tbh]
\begin{centering}
\includegraphics[scale=0.7]{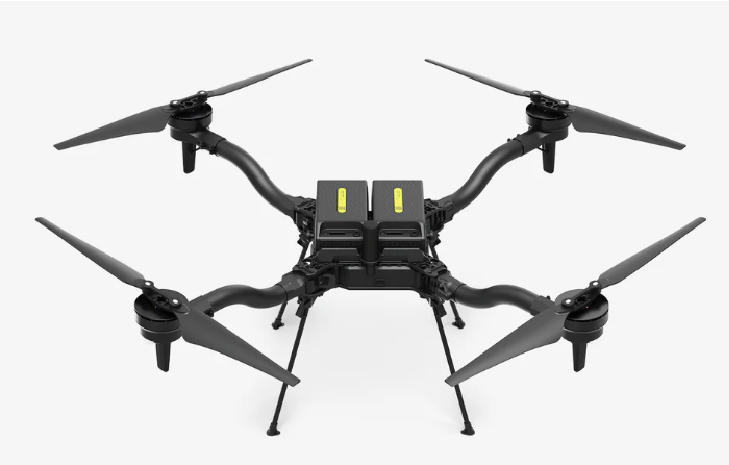}
\par\end{centering}
\caption{\label{fig: astro}The Freefly Astro quadrotor (image courtesy of
Freefly Systems, https://freeflysystems.com/astro).}
\end{figure}

Onboard state estimation is provided by the PX4 EKF2 fusion algorithm,
which combines measurements from GPS (L1/L2 bands supporting GPS,
GLONASS, Beidou, and Galileo), optical flow, lidar, and barometer
sensors. This multi-sensor fusion provides global position, altitude
above mean sea level (AMSL), velocity estimates, and height above
ground.

The LyAT controller is implemented in Python using PyTorch and integrated
into a ROS2 framework. The ROS2 control loop operates at 50 Hz, though
computational demands of the transformer architecture result in control
commands being transmitted to the flight controller at approximately
20 Hz. Velocity commands are communicated to the PX4 autopilot via
the MAVROS package, which converts ROS2 topics to MAVLink protocol
messages. A maximum velocity saturation of $\text{VEL}_{\max}=1.8$
m/s is enforced on all control commands to ensure safe operation within
the experimental flight zone.

\subsection{Experimental Settings}

The quadrotor is set to autonomously track a figure-8 desired trajectory
modeled as
\[
x_{d}=\left[\begin{array}{c}
a\sin\left(\omega t\right)\\
b\sin\left(2\omega t\right)\\
h\\
a\omega\cos\left(\omega t\right)\\
2b\omega\cos\left(2\omega t\right)\\
0
\end{array}\right],
\]
where the height is $h=2.5\,\text{m}$, the angular frequency is $\omega=0.15\,\text{rad/sec}$,
$a=7.5\,\text{m}$, and $b=3\,\text{m}$. The initial conditions are
as $x\left(0\right)=\left[0,0,2.5,0,0,0\right]^{\top}.$ The LyAT
architecture is configured with $N=1$ encoder and decoder layers,
with all multi-head attention mechanisms (self-attention, cross-attention,
and masked self-attention) utilizing $H=3$ attention heads. The feedforward
dimension is set to $d_{f}=5$, and the historical window size is
$\boldsymbol{\tau}=20$ data points. During the initial transient
phase before $\boldsymbol{\tau}=20$ historical data points are available,
the decoder input is populated with random Gaussian noise scaled by
0.1, which are sequentially replaced with actual network outputs as
they become available. Similarly, the encoder input uses zero-padding
for unavailable historical states. The LayerNorm parameters are $\gamma_{f}=\gamma_{1,1}=0.8$,
$\gamma_{1,2}=\gamma_{1,3}=\gamma_{F}=0.7$, and $\beta_{f}=\beta_{1,1}=\beta_{1,2}=\beta_{1,3}=\beta_{F}=0$.
The transformer weights are initialized using Xavier uniform initialization
with gain $0.01$. The adaptation learning rate is selected as $\Gamma=0.02I$,
and the forgetting factor is set as $\sigma=10^{-6}$. The parameter
bound and the control gain are selected as $\overline{\theta}=10$
and $k_{e}=0.8$, respectively.

\subsection{Results and Discussion}

\begin{figure}[tbh]
\begin{centering}
\includegraphics[scale=0.25]{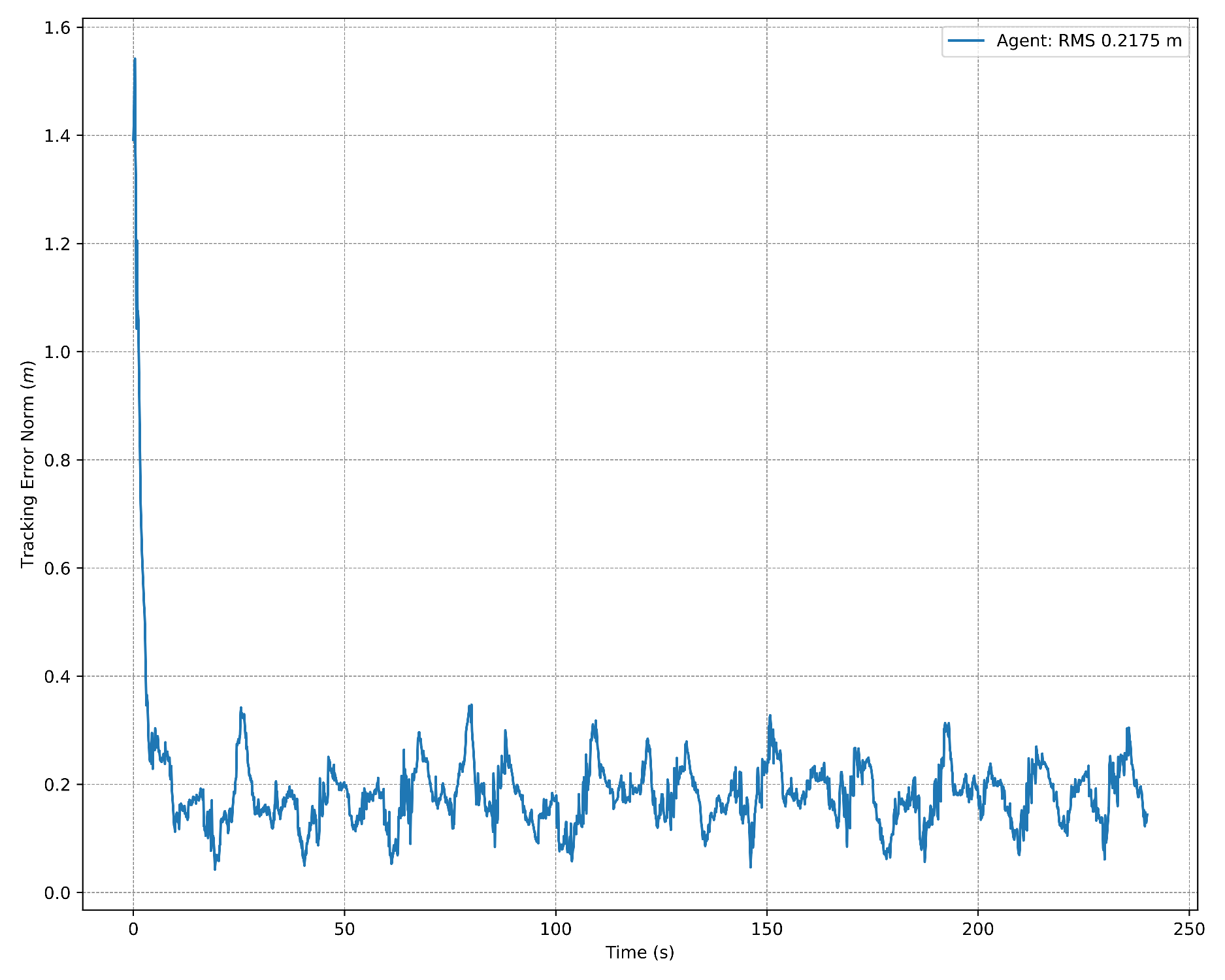}
\par\end{centering}
\caption{\label{fig: RMS of tracking error}Evolution of the RMS of the tracking
error over 240 seconds.}
\end{figure}

The RMS tracking error in Figure \ref{fig: RMS of tracking error}
exhibits an initial transient peaking at approximately 1.5 meters,
which is due to the initial condition error. After approximately 10
seconds, the error converges to a steady-state ultimate bound with
typical amplitudes between 0.1 to 0.3 meters. The persistent oscillations
are potentially attributed to environmental disturbances, sensor noise
and bias, and uncertain dynamics. The RMS tracking error across the
entire 240-second flight is 0.2175 meters. Notably, the error remains
uniformly bounded throughout the experiment, validating the theoretical
UUB-p guarantee from Theorem \ref{thm: main stability theorem}.

\begin{figure}[tbh]
\begin{centering}
\includegraphics[viewport=75bp 0bp 810bp 720bp,clip,scale=0.35]{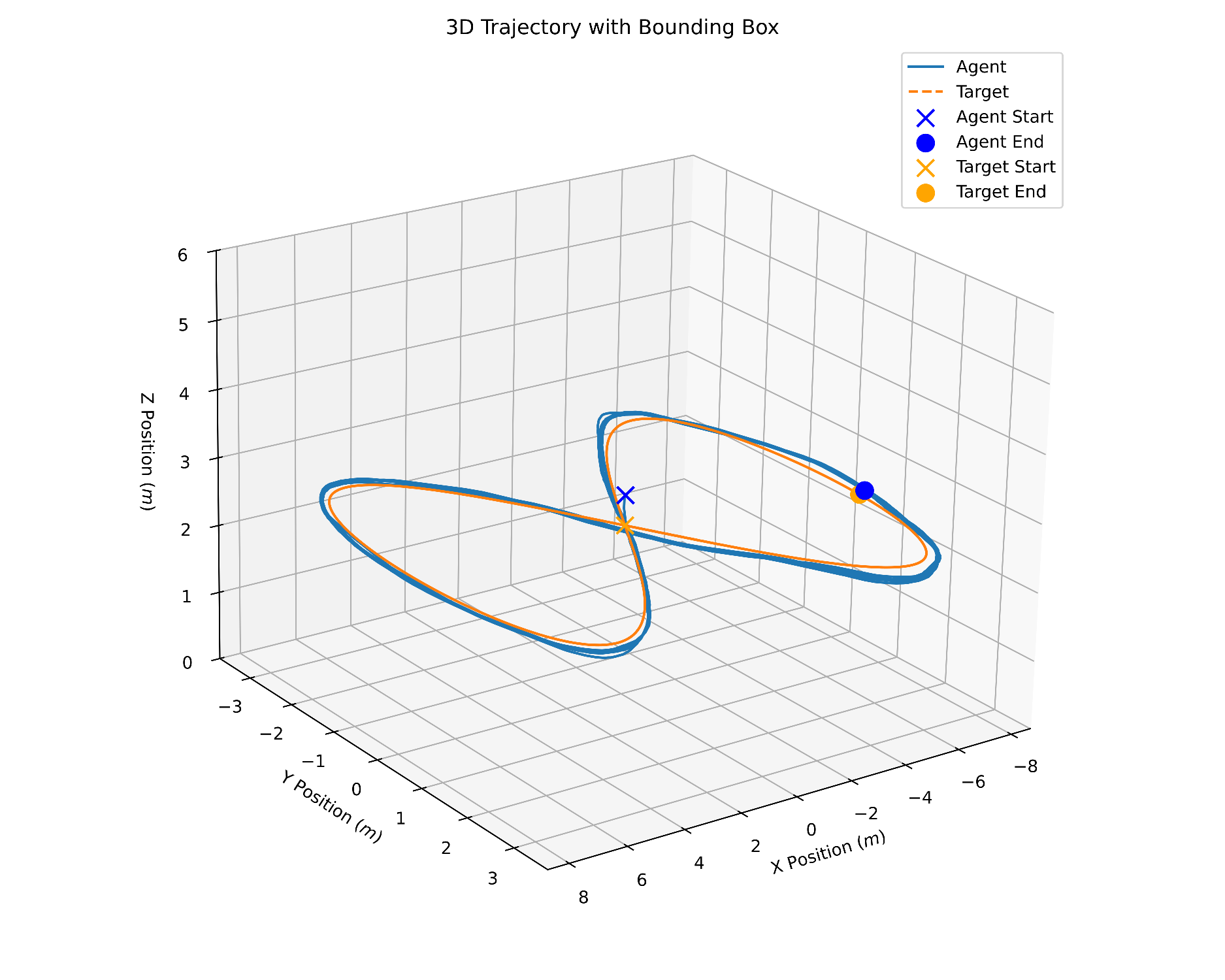}
\par\end{centering}
\caption{\label{fig: 3d schematic of the trajectory}Three-dimensional schematic
of the trajectory tracking.}
\end{figure}

The three-dimensional trajectory visualization in Figure \ref{fig: 3d schematic of the trajectory}
confirms the successful tracking of the figure-8 desired trajectory
over 5.7 cycles. The quadrotor maintains close proximity to the desired
trajectory throughout the maneuver, with slight deviations primarily
visible during high-curvature portions at the figure-8 crossover points.
The consistent altitude maintenance around 2.5m demonstrates effective
compensation of both drift and diffusion uncertainties.

\begin{figure}[tbh]
\begin{centering}
\includegraphics[scale=0.25]{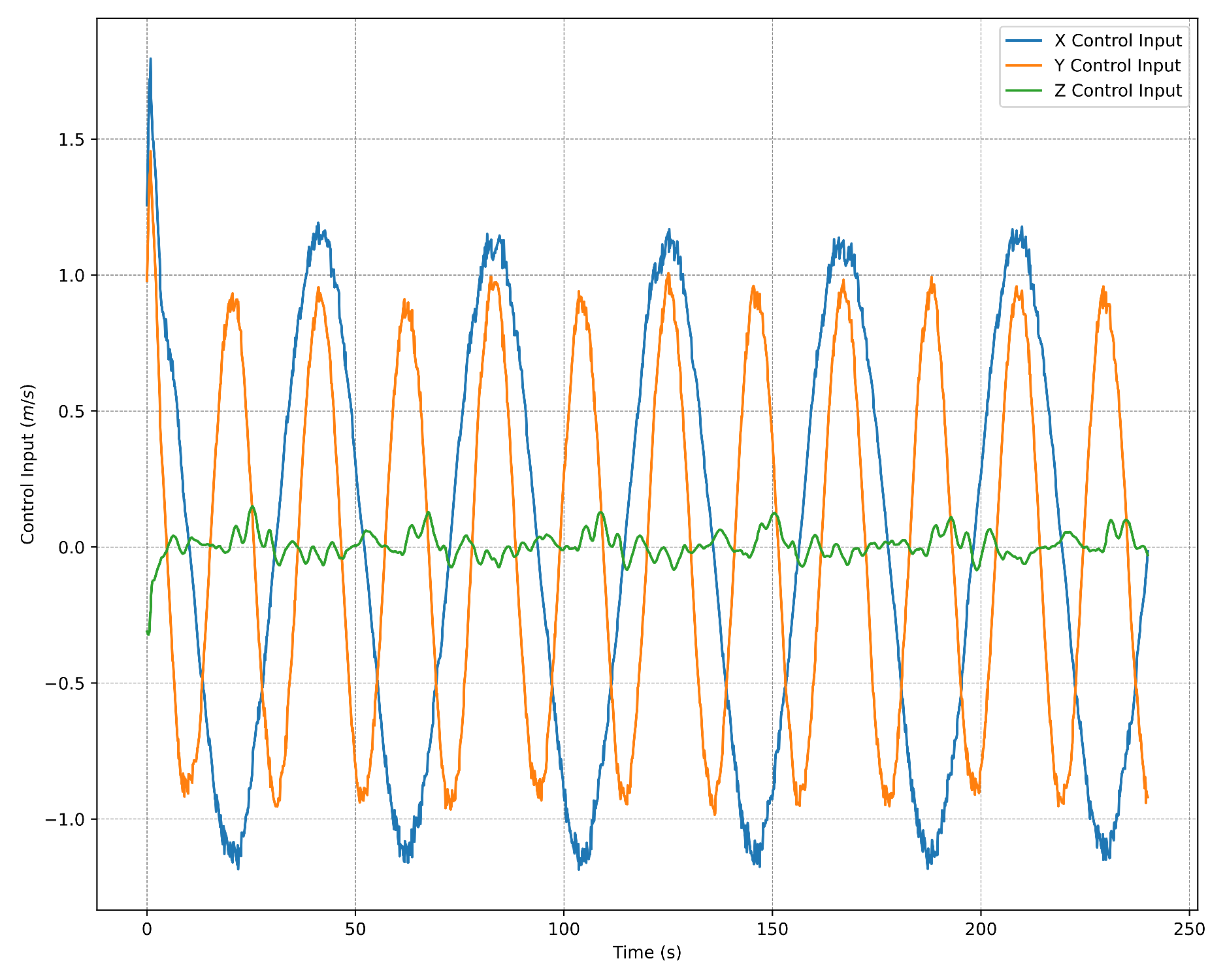}
\par\end{centering}
\caption{\label{fig: control input}Evolution of control input over 240 seconds.}
\end{figure}

Figure \ref{fig: control input} illustrates the control effort across
all three spatial dimensions. To track the horizontal figure-8 motion,
the $X$ and $Y$ channels show larger oscillations with respective
peak magnitudes of approximately 1.2 m/s and 1 m/s, while the Z-channel
requires minimal actuation of \textpm 0.2 m/s to maintain constant
altitude. The initial control spike corresponds to the aggressive
corrective action in response to the large initial tracking error.
Following the transient, the control signals remain smooth and continuous,
demonstrating the LyAT's ability to generate implementable commands
without chattering or saturation.

These experimental results validate that the proposed LyAT controller
successfully adapts to system uncertainties in real-time without requiring
offline training, maintains bounded tracking error with RMS performance
of 0.2175 meters, and generates smooth, implementable control signals
suitable for physical quadrotor deployment. The rapid convergence
and sustained performance over the full 240-second flight validate
the effectiveness of the Lyapunov-based adaptation mechanism in compensating
for both drift and diffusion uncertainties under real outdoor flight
conditions.

\section{Conclusions}

This paper presented a Lyapunov-based adaptive transformer controller
for adaptive control of stochastic nonlinear systems with unstructured
uncertainties. The developed transformer architecture features analytically-derived
weight adaptation laws that enable real-time parameter updates without
offline training requirements. A constructive Lyapunov-based stability
analysis guarantees probabilistic uniform ultimate boundedness of
tracking and parameter estimation errors. The unified neural network
architecture simultaneously compensates for drift and diffusion uncertainties,
significantly reducing computational complexity compared to prior
methods requiring multiple separate networks \cite{Akbari.Nino.ea2024}.
Experimental validation on a Freefly Astro quadrotor tracking a figure-8
trajectory demonstrated rapid convergence to an ultimate and sustained
performance with RMS tracking error of 0.2175 meters. The results
confirm the LyAT's real-time adaptation capabilities and theoretical
performance guarantees under real-world flight conditions with environmental
disturbances.

\bibliographystyle{ieeetr}
\bibliography{encr,master,ncr}

\end{document}